\documentclass[useAMS,usenatbib]{elsarticle} 

\usepackage{amssymb,amsmath}
\usepackage{psfig,epsfig,amssymb,alltt}
\usepackage{abbrevjournaux}
\usepackage{latexsym}
\usepackage{graphicx}
\graphicspath{{PS/}}
\usepackage{psfig}
\psfigurepath{./PS:.}
\usepackage[english]{babel}
\usepackage{indentfirst}

\title{Towards a fast, model-independent Cosmic Microwave Background bispectrum estimator}
\author[sp]{S.~Pires}
\author[spl]{S. Plaszczynski}
\author[spl]{A. Lavabre}
\address[sp]{AIM, CEA/DSM-CNRS-Universite Paris Diderot, IRFU/SEDI-SAP, Service d'Astrophysique, \\CEA Saclay, Orme des Merisiers, 91191 Gif-sur-Yvette, France\\}
\address[spl]{Laboratoire de l'Acc\'el\'erateur Lin\'eaire (LAL),\\ CNRS: UMR8607, IN2P3, Universit\'e Paris-Sud, Orsay,  France}
\begin{document}

\begin{abstract}
The measurements of the statistical properties of the Cosmic Microwave Background (CMB) fluctuations enable us to probe the physics of the very early Universe especially at the epoch of inflation. 
A particular interest  lays on the detection of the non-Gaussianity of the CMB as it can constrain the current proposed models of inflation and structure formation, or possibly point out new models. 
The current approach to measure the degree of non-Gaussianity of the CMB is to estimate a single parameter 
$f_{\rm NL}^{\rm local}$ which is highly model-dependent.
 The bispectrum is a natural and widely studied tool for measuring the non-Gaussianity in a model-independent way.
This paper sets the grounds for a full CMB bispectrum estimator
based on the decomposition of the sphere onto projected patches. The mean bispectrum estimated this way can be calculated quickly and is model-independent. This approach is very flexible, allowing exclusion of some patches in the processing or consideration of just a specific region of the sphere.
\end{abstract}


\maketitle

\section{Introduction}
\label{sect_intro}
Inflation is the currently favored theory of early Universe which predicts an early short period of rapid expansion and explains the origin of primordial perturbations. 
Many models of inflation predict weakly non-Gaussian primordial curvature perturbations and the primordial curvature perturbation $\Phi(x)$ is parametrized using the local model \citep{bispec:komatsu01} as follows:
\begin{equation}
\Phi(x)  = \Phi_ {\rm L}(x)+f^{local}_{\rm NL} (\Phi_{\rm L}^2(x)-\langle \Phi_{\rm L}^2(x) \rangle),
\label{local}
\end{equation}
where $f^{\rm local}_{\rm NL}$ is the non-linear coupling constant in the local model.

Due to its simplicity the local model is highly favored especially because all the higher-order moments are determined in terms of this  $f^{\rm local}_{\rm NL}$ parameter. 
Most of models of inflation only predict a value for $f^{\rm local}_{\rm NL}$ to characterize the non-Gaussianity of the CMB. 
However, there exists other models of inflation that predict different types of deviations from Gaussianity, detailed calculation of which have been investigated in \cite{bispec:creminelli06,bispec:fergusson09}. \\

The observed CMB temperature fluctuations $\Delta T/T$ are related to the primordial curvature perturbation $\Phi(x)$ through the following non-linear relation \citep{bispec:komatsu02b}: 
\begin{equation}
\frac{\Delta T}{T} \approx g_{\rm T}(\Phi(x)),
\label{perturbation}
\end{equation}
where $g_{\rm T}$ is the radiation transfer function.
On small scales, this function is very complex and it is evaluated numerically by solving the Boltzmann transport equation. On very large scales, in the Sachs-Wolfe regime ($l < 10$), this relationship simplifies to $\frac{\Delta T}{T} = \frac{\Phi(x)}{3}$. 
A number of theories of inflation have been proposed that make different predictions about the CMB  fluctuations. Measurements of the statistical properties of the CMB are a direct test of inflation which can help to rule out the many models of inflation that have been proposed.

The observed CMB anisotropies $\Delta T/T$ can be expanded onto spherical harmonics:
\begin{equation}
\Delta T/T = \sum_{lm} a_{lm} Y_{lm}.
\label{spherical}
\end{equation}
If the CMB is Gaussian, it is fully described  by its angular power spectrum:
\begin{equation}
C_l = \frac{1}{2l+1}\sum_{m=-l}^{+l}|a_{lm}|^2.
\label{powerspec}
\end{equation}

But, as explained previously, many realistic models predict deviations from Gaussianity. 
Even if, the level of non-Gaussianity is predicted to be very small in single field slow-roll inflation model, there is a large class of more general models that  predict a substantially high level of primordial non-Gaussianity. If these models are true, the power spectrum provides a limited insight to the physics of the very early Universe and higher-order estimators are needed to probe the CMB non-Gaussianity. The bispectrum is a natural model-independent approach to probe the small departure from Gaussianity that could originate during inflation. The CMB angular bispectrum may be calculated from product of three spherical harmonic coefficients of the CMB temperature field. For Gaussian fields, the expectation value  is exactly zero. Given statistical isotropy of the universe, the angular bispectrum $B_{l_1 l_2 l_3}$ is given by (Komatsu et al, 2001):
\begin{eqnarray}
B_{l_1 l_2 l_3} =  \sum_{m_1, m_2, m_3} \left(\begin{array}{ccc} l_1 & l_2 & l_3 \\ m_1 & m_2 & m_3 \end{array} \right) a_{l_1 m_1}a_{l_2 m_2}a_{l_3 m_3},
\label{bispectrum}
\end{eqnarray}
where the matrix denotes the Wigner-3j symbol, $B_{l_1 l_2 l_3}$ satisfy the triangle condition: $|l_1 - l_2| \le l_3 \le l_1 + l_2$ and the parity invariance: $l_1+l_2+l_3=$ even. Then, a non-zero bispectrum is a signature of a more complicated inflationary period than a simple inflation slow-roll model predicts.

However, computation of a full bispectrum is very time-consuming and it is usually assumed that the signal is too weak for each of the multipoles to be measured individually. Instead a least squares fit to compare the observed bispectrum with a particular (separable) theoretical bispectrum is used. Thus, most non-Gaussianity studies focus on estimating the $f^{\rm local}_{\rm NL}$ parameter because the bispectrum is fully specified by this parameter in the case of a local model (\ref{local}). A fast estimator for $f^{\rm local}_{\rm NL}$ has been developed by \cite{bispec:komatsu05} and improved by \cite{bispec:creminelli06}:

\begin{equation}
f^{\rm local}_{\rm NL} \simeq \left[\sum^{l_{\rm max}}_{l_1 \le l_2 \le l_3} \frac{(\mathcal{B}^{\rm prim}_{l_1 l_2 l_3})^2}{C_{l_1}C_{l_2}C_{l_3}}\right]^{-1} S_{\rm prim},
\label{fnl}
\end{equation}
where $C_l$ is the theoretical power spectrum and $\mathcal{B}^{\rm prim}_{l_1 l_2 l_3}$ is the theoretical bispectrum \citep{bispec:komatsu01} for $f^{\rm local}_{\rm NL}=1$. The statistics $S_{\rm prim}$ defined in \citep{bispec:komatsu03} has only a complexity of $\mathcal{O}(N^{3/2})$ whereas the full bispectrum analysis is $\mathcal{O}(N^{5/2})$.
A detection of $f^{\rm local}_{\rm NL} > 10$ will rule out most of the existing inflation models.\\

However, an expression like equation (\ref{local}) is not general and there are many other inflationary models that predict different types of deviations from Gaussianity. Other models have been investigated and detailed calculations of other form of non-Gaussianity have been carried out (see for example \cite{bispec:creminelli06,bispec:fergusson09}). However, all these estimators are
highly model-dependent and may not be able to constrain, many of the non-Gaussiannity signatures. In this paper, we will investigate a more general and model-independent approach to probe the CMB non-Gaussiannity by measuring the full bispectrum. If the non-Gaussianities are of the local type, the bispectrum will reach a maximum in case of squeezed configurations (i.e. one wave vector is much smaller than the other two). \\


Full bispectrum estimation on spherical data using the spherical harmonics definition (\ref{bispectrum}) is very time consuming. This paper tries to tackle this problem by introducing a promising approach for accelerating the calculation of the fully general bispectrum. This method is based on the projection of the sphere onto small-field projected maps for which a Fourier decomposition is used to estimate the bispectrum. A mean bispectrum estimator can then be obtained by combining results from all the patches. This paper sets the grounds for this method however for full optimization of the estimator, we need more accurate non-Gaussian simulations which are not yet available and an analytical prediction of a full bispectrum for a given level of non-Gaussianity which is very complex to code up.

The outline of the paper is as follows. In \S \ref{sect_method}, the main issues of the method are pointed out and the method is optimized to do power spectrum estimation. In \S \ref{sect_application} the same method is used to accelerate the full bispectrum estimation. Some adaptations of the method are required to optimize the bispectrum estimation.

\section{Decomposition of the sphere onto rectangular Cartesian maps}
\label{sect_method}

In this section, we will describe a method to speed up the spectral analysis by decomposing the sphere onto patches. Such method is similar to the Welch's method \citep{stft:oppenheim75,stft:welch67} that is commonly used to reduce the variance in spectral analysis of 1D data. For a sphere, the size and the repartition of the patches have to be decided. Then, the pixels of the patches have to be projected onto rectangular Cartesian maps. A spectral estimator is then obtained by averaging the result of the spectral analysis on each rectangular Cartesian map. Such methods suffer from the limited size of the projected patches which introduces a bias. In a future paper, we will consider multi-taper techniques \cite{multitaper:das09} that consist in averaging over different tapers using the full data. These methods reduce the bias since the data length is not shortened. 

In this section, we will follow the basic approach that consists in decomposing the sphere onto patches.
A number of issues has to be solved before having an optimal estimator.

\subsection{Some issues of the method}
\label{sect_method_issue}
The first issue to be tackled is to find an optimal tessellation of the sphere. To facilitate the post-processing, especially the FFT required to do spectral analysis, the patches ought to be rectangular and preferably have the same orientation. However, no regular tessellation of the sphere by rectangles exists. Instead, a pseudo regular tessellation with rectangles can be obtained by allowing the patches to overlap. 
The overlapping also improves the power spectrum estimation : according to the Welch's overlapped segment method \citep{stft:welch67,stft:oppenheim75}, estimating the power spectrum of a signal by splitting this one into overlapping windowed patches reduces the noise in the estimated power spectra by reducing the frequency resolution.
There still remains the size and position of the patches to be determined.
To reduce the distortions, the size of the patches should depend on the map projection and is a compromise between the distortions introduced by the projection and the window function effects. \\

The {\bf effect of the window function} on the final power spectrum is another issue to take into account. The limited size of the patches forces us to use window functions. Thus, instead of analyzing the signal $s(x,y)$, we will be analyzing the truncated signal: $s_h(x,y) = s(x,y) h(x,y)$. In the frequency domain, we obtain the following convolution product $S_h(u,v) = S(u,v)*H(u,v)$ where $H(u,v)$ is the Fourier transform of the window function. By default the window function is a rectangular window, constant inside the patch and zero elsewhere (see the left panel of the Fig.\ref{window}). But, it appears that a window function with better frequency response has to be used. The ideal window function is one whose frequency response is a Dirac delta function. It corresponds to an infinite rectangular window function which is impossible in practice. Instead, the frequency response normally consists of a main lobe and side lobes (see the right panel of the Fig.\ref{window}). To be close to a Dirac function, the main lobe ought to be the highest and the narrowest to increase the frequency resolution and the side lobes have to be the lowest to limit the mode-to-mode interaction. The rectangular window function is usually not recommended because of its significant sides lobes.\\


An additional issue comes from the {\bf projection effects}. The method that we have developed is based on the decomposition of the sphere onto a few rectangular Cartesian maps. It assumes to do pixel projections from the HEALPix map to rectangular Cartesian maps. No matter how sophisticated the projection process will be, distortions are inherent in flattening the sphere. Some classes of map projections maintain areas, and others preserve local shapes, distances, and/or directions... No projection, however, can preserve all these characteristics. Choosing a projection thus always requires compromising accuracy in some way, and that is one reason why so many different map projections have been developed. \\

Whatever the projection, if we want to keep this projection exact, we will have to deal with rectangular Cartesian maps having a {\bf non-regular grid}. This will be a problem especially for the power spectrum and bispectrum estimation because we need to estimate the Fourier coefficients from a signal $f(x, y)$ at arbitrary nodes ($x, y$). \\

Assuming the signal $f$ is a periodic function, it can be decomposed into Fourier series as follows:
\begin{equation}
f(x, y) = \sum_{n_1=-\infty}^{+\infty} \sum_{n_2=-\infty}^{+\infty} C_{n_1,n_2}(f) e^{2 \pi i \frac{n_1}{T}x} e^{2 \pi i \frac{n_2}{T}y}
\label{series}
\end{equation}

The approximation that has been used in this paper  consists in finding the coefficients $C_{n_1,n_2}(f)$ of the Fourier series (\ref{series}).
The problem can be solved efficiently by noting the system matrix is a Toeplitz matrix i.e. a diagonal-constant matrix \citep[see][ for more details about the use of Toeplitz matrices in irregular sampling problems]{toeplitz:Feichtinger95}

\subsection{Optimization of the method for power spectrum estimation}
\label{sect_power}
As discussed in the previous section, a number of issues has to be solved in order to use the previous method for spectral estimation. In this section, the method is optimized to perform power spectrum estimation. For this purpose, we have generated simulations of the full sky CMB temperature as realizations of a random Gaussian field with a prescribed power spectrum. The cosmology adopted in these simulations is consistent with the WMAP parameters. We use the HEALPix pixelisation as WMAP and Planck missions  with a resolution parameter of $n_{\rm side}=1024$. 

\subsubsection{The tiling of the sphere}
\label{sect_power_method_tiling}
Following the ideas discussed in \S \ref{sect_method}, several decompositions have been tested. \\

\begin{figure}[htp!]
\centerline{
\includegraphics[width=8.cm, height=4.2cm]{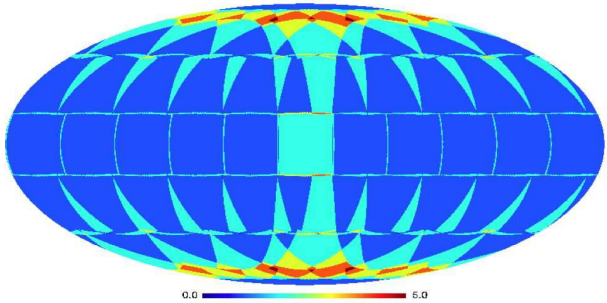}
}
\caption{Sphere tiling using an equi-latitude decomposition}
\label{tiling1}
\end{figure}

An interesting tiling consists in decomposing the sphere onto rectangular patches distributed into lines of same latitude (see Fig.\ref{tiling1}).
This decomposition introduces a substantial overlapping at the pole. For power spectrum estimation, this effect can be neglected by assuming the CMB field is isotropic. But this decomposition should not be used to detect non-Gaussianity because some types of non-Gaussianity can produce localized hot spots or other structures.\\

\begin{figure}[htp!]
\centerline{
\includegraphics[width=8.cm, height=4.2cm]{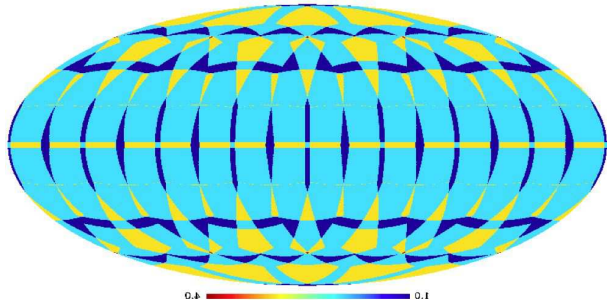}
}
\caption{Sphere tiling using the HEALPix centers of a lower resolution.}
\label{tiling2}
\end{figure}

Another tiling consists in decomposing the sphere onto rectangular patches located at the HEALPix centers of a lower resolution (see Fig.\ref{tiling2}). This decomposition insures a good repartition of the patches in the sky and should be preferred for bispectral estimation.

Once, we have selected the optimal tiling of the sphere, the pixels of the patches are projected onto rectangular Cartesian maps using a projection that will be described in \S \ref{sect_power_method_proj}. To perform the decomposition, the size of the patches has to be fixed in such a way that keeps the induced distortions to minimum. To reduce the spectral leakage, the size of the patch has to be increased to narrow the main lobe of the frequency response. But increasing the size of the patch will increase the distortions effects due to projection effects.

\begin{figure}[htp!]
\begin{center}
\begin{tabular}{rcc|}
\includegraphics[width=6.2cm, height=4.2cm]{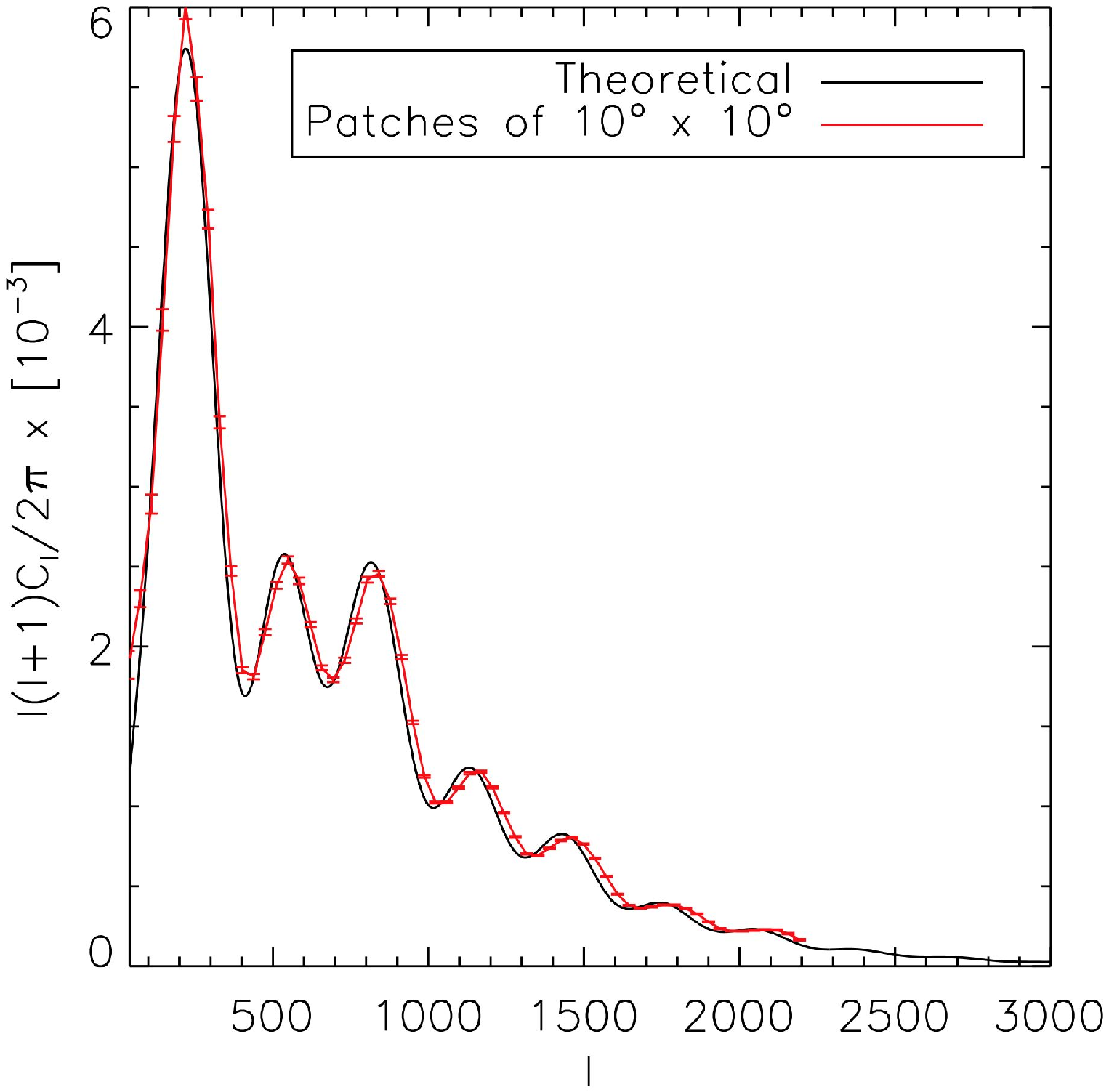}&
~~~\includegraphics[width=5.cm, height=4.2cm]{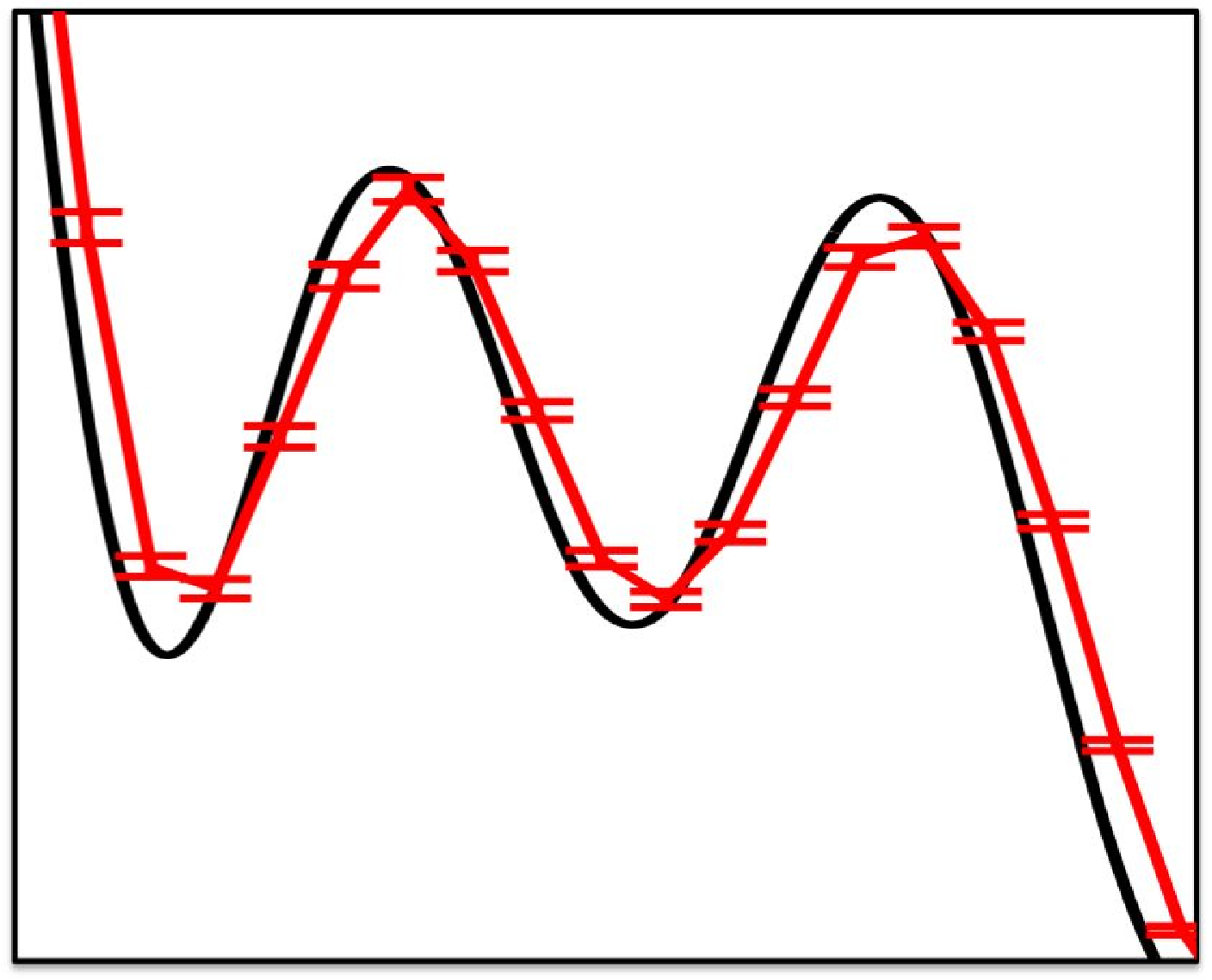}\\
\includegraphics[width=6.2cm, height=4.2cm]{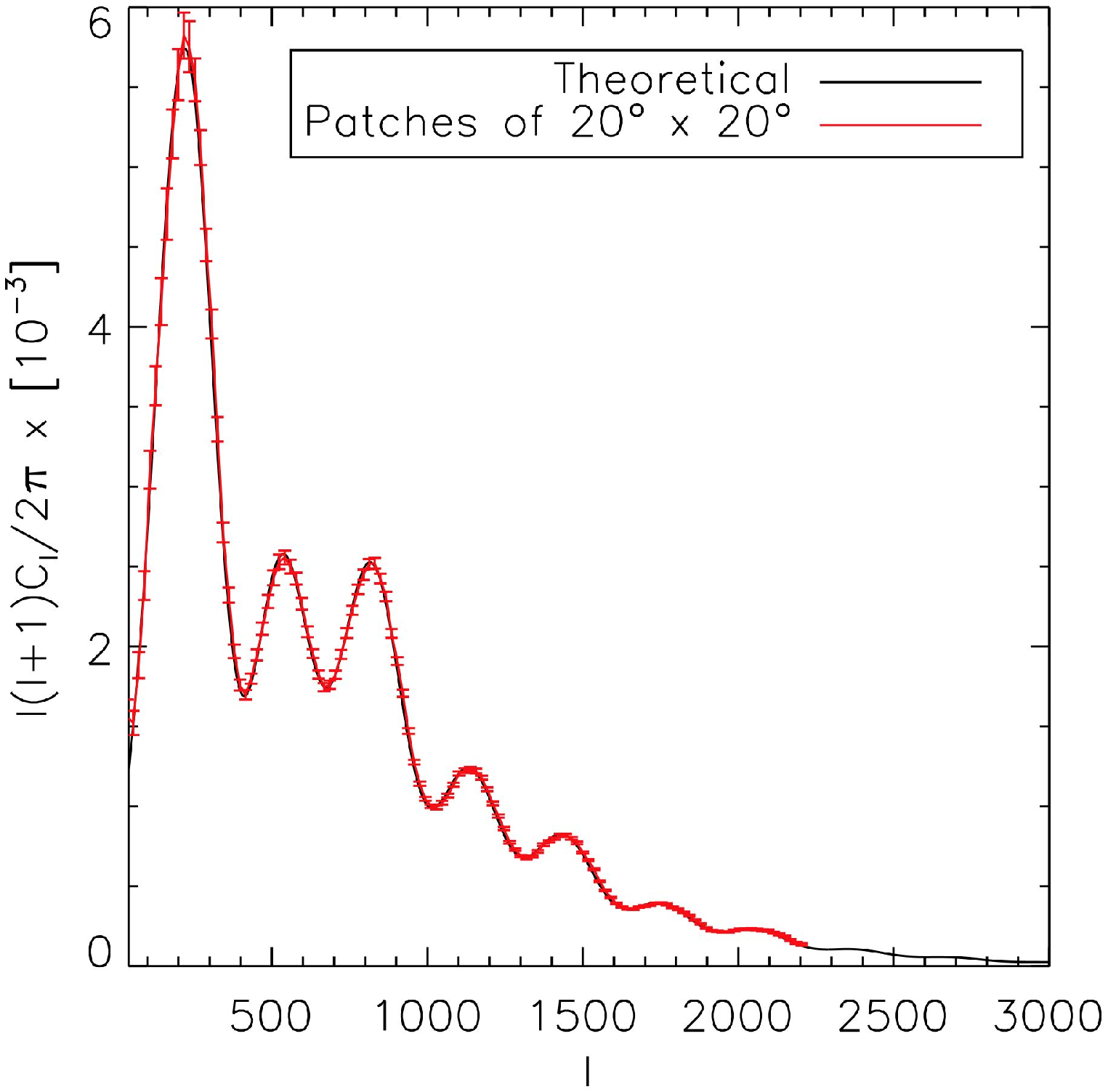}&
~~~\includegraphics[width=5.cm, height=4.2cm]{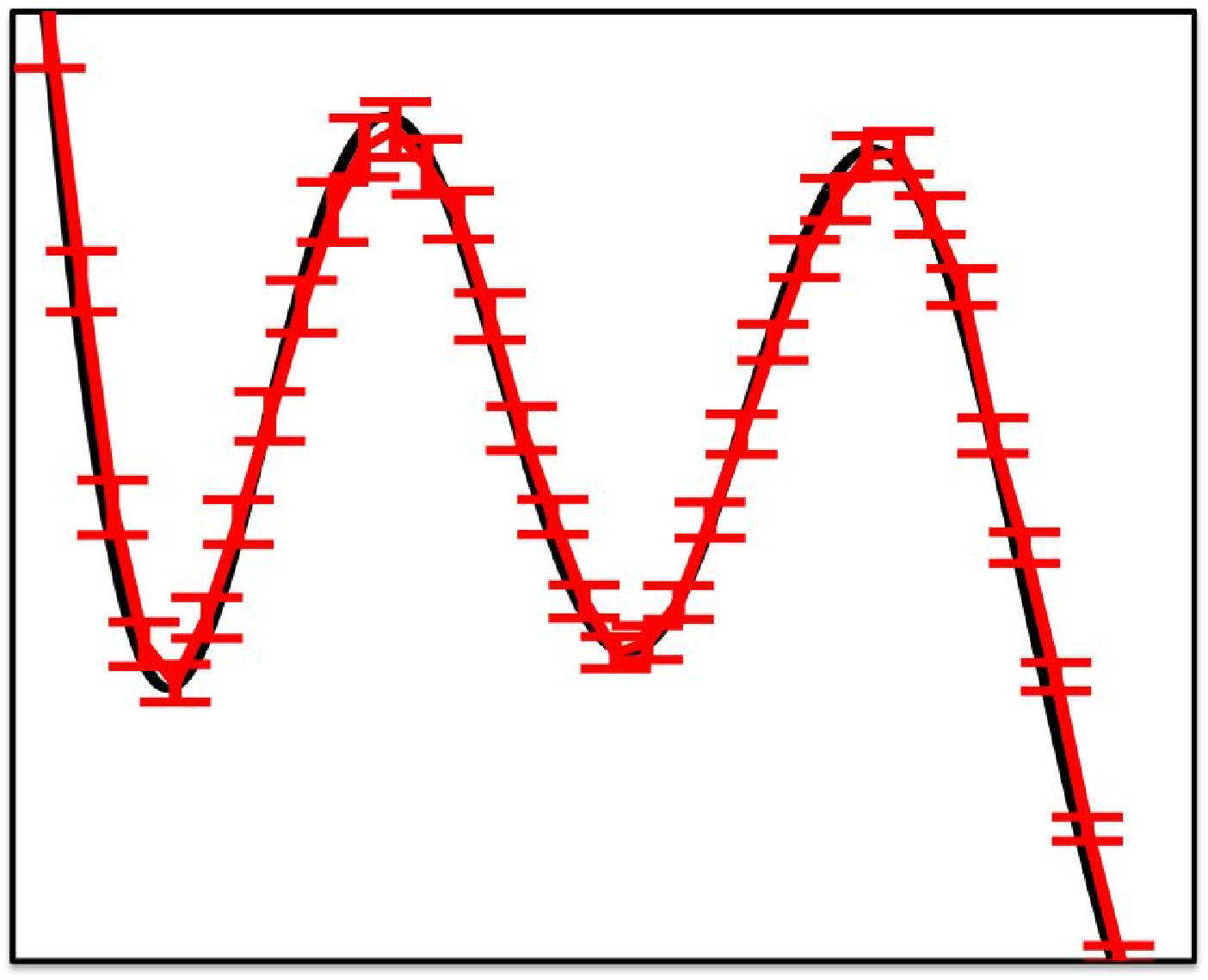}\\
\includegraphics[width=6.2cm, height=4.2cm]{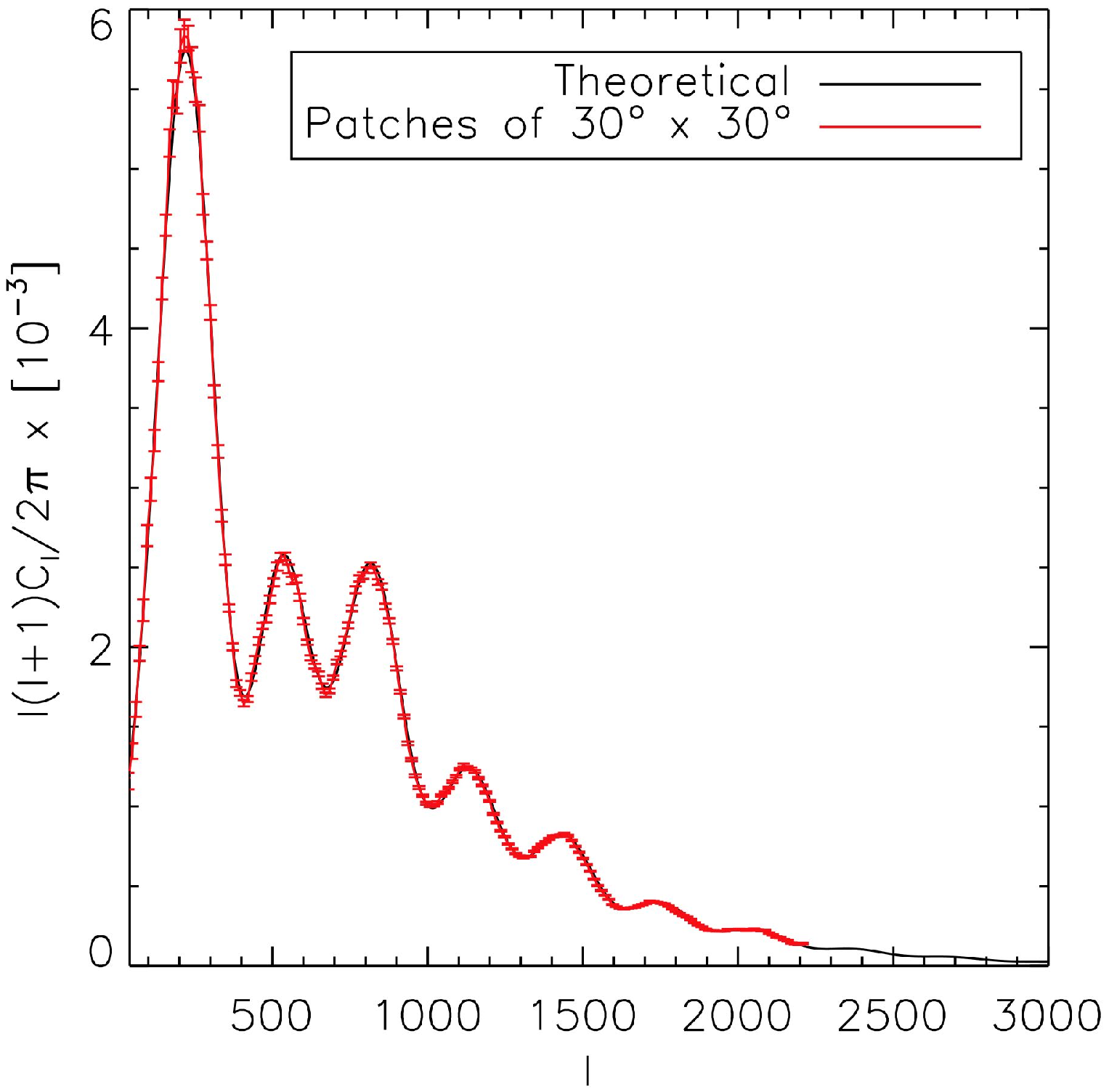}&
~~~\includegraphics[width=5.cm, height=4.2cm]{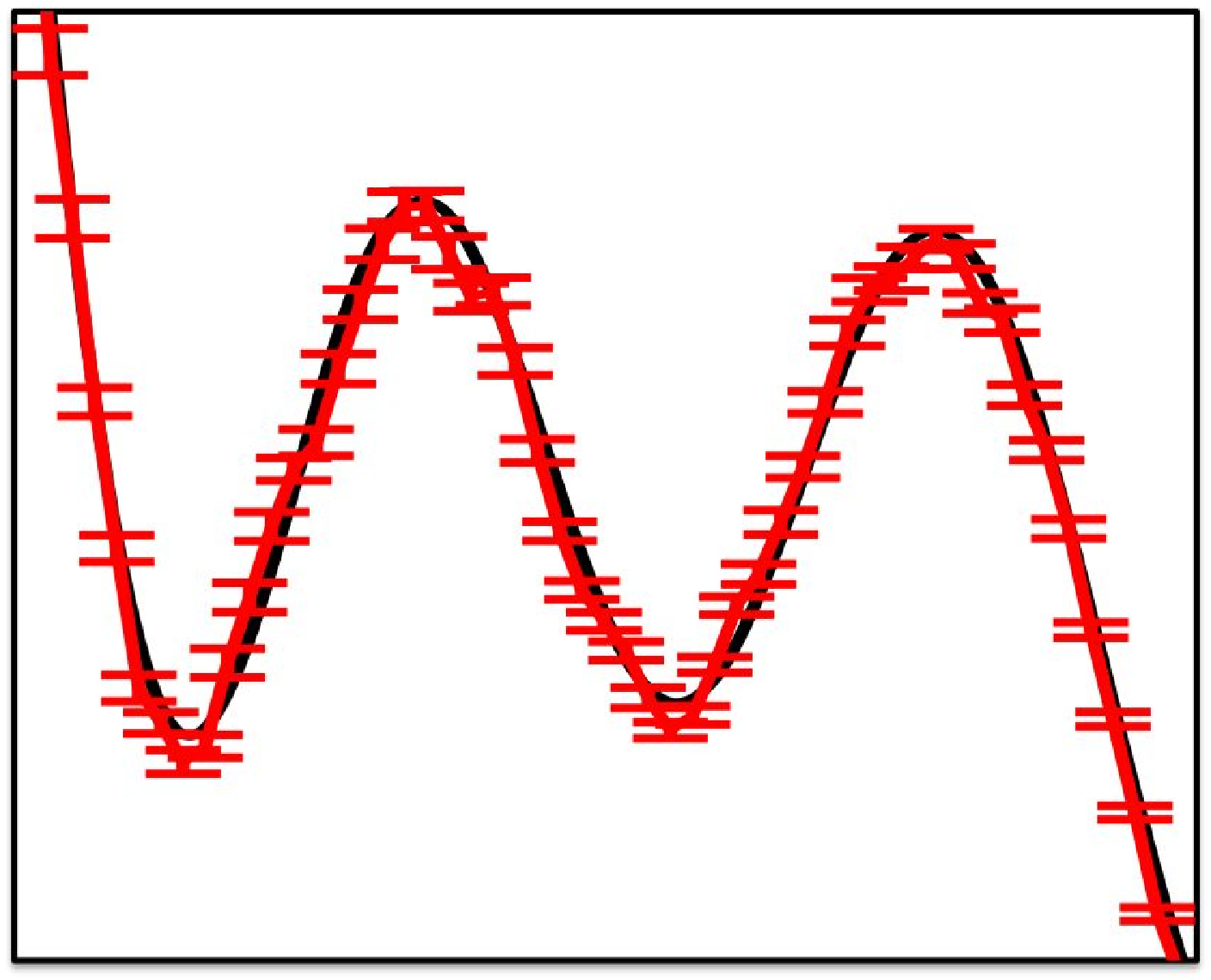}\\
\end{tabular}
\caption{{\bf The effect of the size of the patches in the power spectrum estimation:} a power spectrum (with error bars) has been estimated using three different size of patches (Note: a Hann window function is used to reduce the spectral leakage): 10$^{\circ}$x10$^{\circ}$ (top left), 20$^{\circ}$x20$^{\circ}$ (middle left) and 30$^{\circ}$x30$^{\circ}$ (bottom left). Then, these three power spectra (red) have been compared to the theoretical power spectrum (black). The right column corresponds to a zoom in on the left power spectrum.}
\label{size}
\end{center}
\end{figure}

In Fig.\ref{size}, the mean power spectrum has been estimated using three different sizes of field: $10^{\circ}$x$10^{\circ}$, $20^{\circ}$x$20^{\circ}$ and 30$^{\circ}$x30$^{\circ}$. A Hann window has also been used to reduce the spectral leakage. Then, these power spectra (red) have been compared to the theoretical power spectrum used to simulate the full-sky CMB (black).  


As expected, an important spectral leakage is observed on the mean power spectrum estimated with patches of 10$^{\circ}$x10$^{\circ}$ (see top panel of Fig.~\ref{size}). With patches of 30$^{\circ}$x30$^{\circ}$  (see bottom panel of Fig.~\ref{size}), the spectral leakage is severely dampened but some distortions are visible at large scales, most certainly due to projection effects. The best result is obtained with a field of $20^{\circ}$x$20^{\circ}$  (see middle panel of Fig.~\ref{size}) which is a good compromise to both minimize the spectral leakage and reduce the distortions introduced by the map projection. 

\subsubsection{Map projection}
\label{sect_power_method_proj}
The map projection introduces distortions of various classes, as it has been discussed in \S \ref{sect_method_issue}. The choice of the map projection has to be done in such a way that it reduces the error in the estimated power spectrum. Two projections have been compared:



\begin{itemize}
\item[1.] The gnomonic projection is constructed by projecting every point of the sphere onto patches from the center of the sphere. 
Assuming the patch is tangent to the point S ($\theta_{\circ}$, $\psi_{\circ}$), the coordinate transformation is the following:
\begin{equation} 
\left(\begin{array}{c} x\\ y \end{array}\right) =\left(   \begin{array}{c} \frac{\cos \theta \sin \theta_{\circ} - \sin \theta \cos \theta_{\circ} \cos(\psi-\psi_{\circ})}{\cos \theta \cos \theta_{\circ} + \sin \theta \sin \theta_{\circ} \cos(\psi - \psi_{\circ})} \\ \frac{\sin \theta \sin(\psi -\psi_{\circ})}{\cos \theta \cos \theta_{\circ} + \sin \theta \sin \theta_{\circ} \cos(\psi - \psi_{\circ})}  \end{array}\right),
\end{equation}
where $\theta$ is the longitude and $\psi$ is the latitude.

\item[2.] The stereographic projection is constructed by projecting every point of the sphere onto patches from the sphere north pole 
 in a plane tangent to the south pole.
 Assuming the patch is tangent to the point S ($\theta_{\circ}$, $\psi_{\circ}$), the coordinate transformation is the following:
\begin{equation} 
\left(\begin{array}{c} x\\ y \end{array}\right) =\left(   \begin{array}{c} \frac{2 (\cos \theta \sin \theta_{\circ} - \sin \theta \cos \theta_{\circ} \cos(\psi-\psi_{\circ}))}{1+ \cos \theta \cos \theta_{\circ} + \sin \theta \sin \theta_{\circ} \cos(\psi - \psi_{\circ})} \\ \frac{2 \sin \theta \sin(\psi -\psi_{\circ})}{1+ \cos \theta \cos \theta_{\circ} + \sin \theta \sin \theta_{\circ} \cos(\psi - \psi_{\circ})}  \end{array}\right).
\end{equation}
\end{itemize}


\begin{figure}[htp!]
\begin{center}
\begin{tabular}{rcc|}
\includegraphics[width=6.2cm, height=4.2cm]{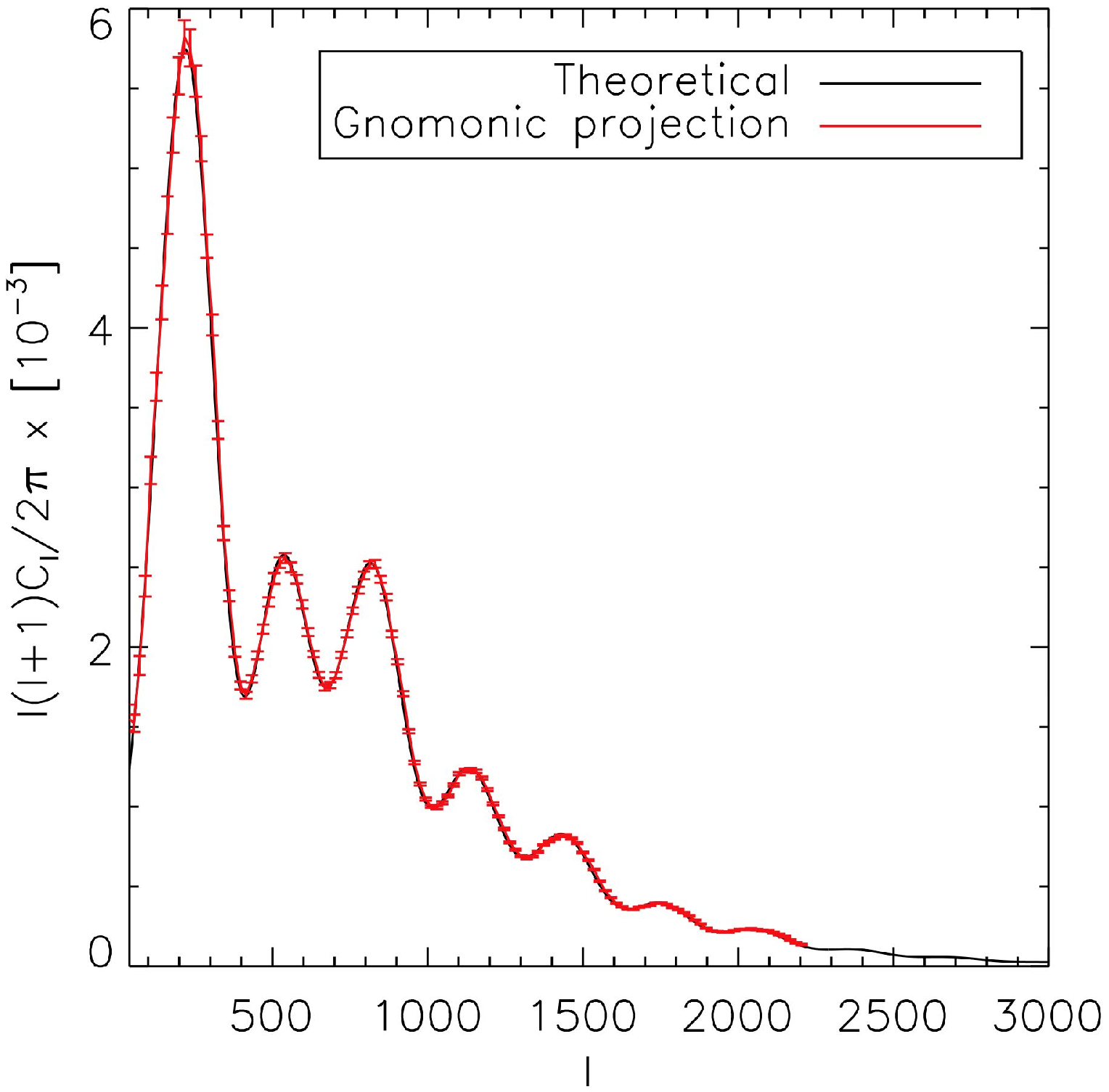}&
~~~\includegraphics[width=5.cm, height=4.2cm]{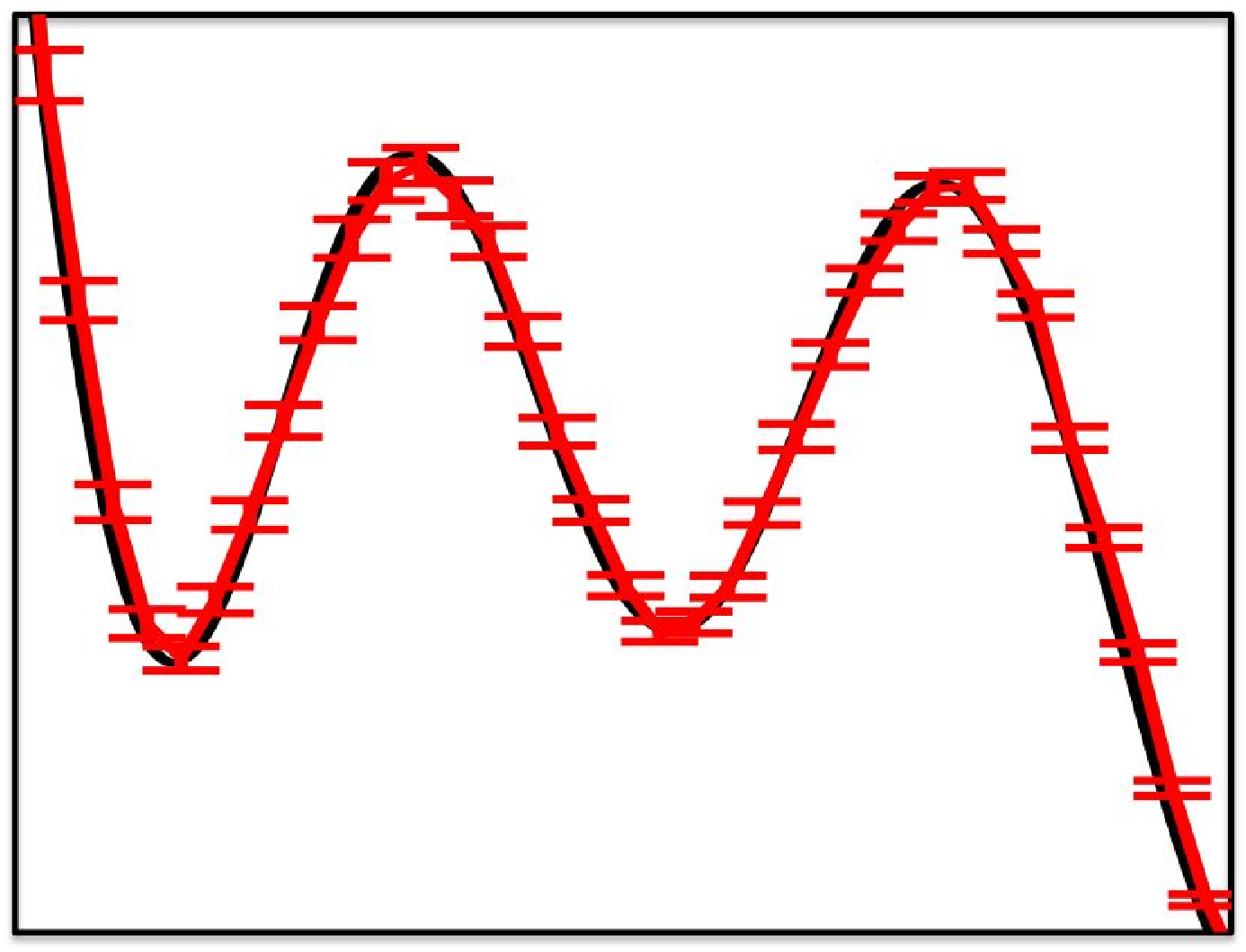}\\
\includegraphics[width=6.2cm, height=4.2cm]{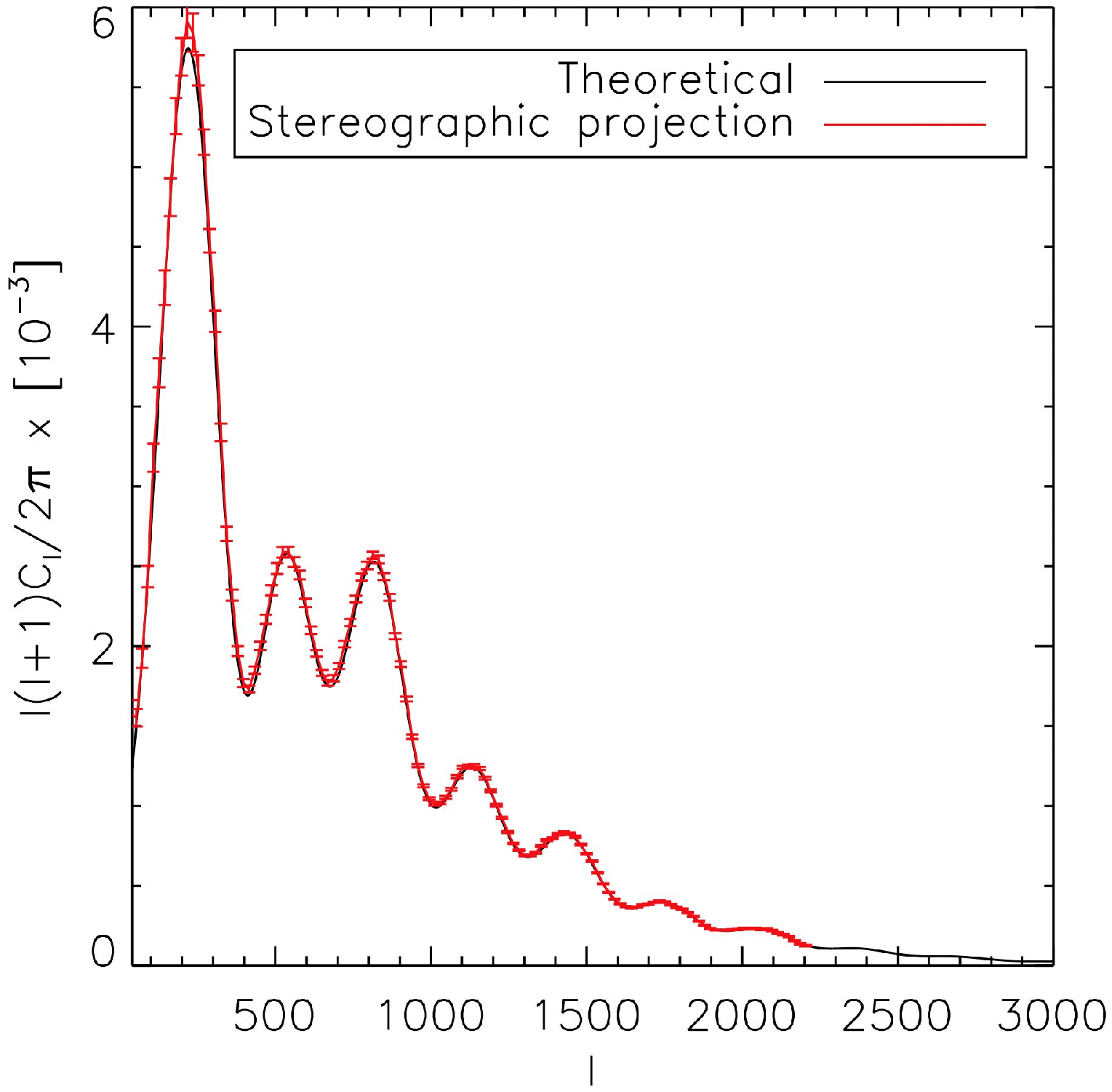}&
~~~\includegraphics[width=5.cm, height=4.2cm]{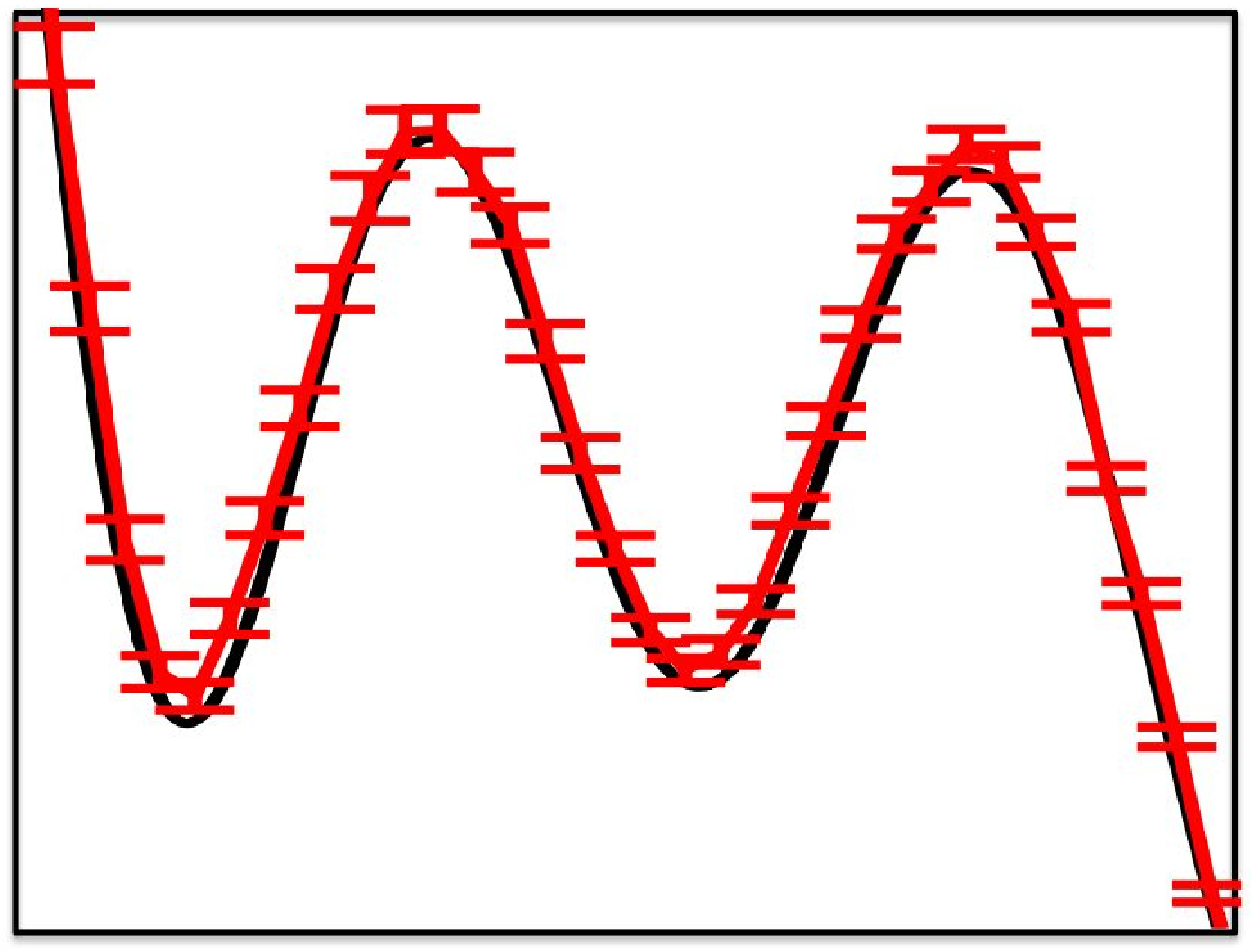}\\
\end{tabular}
\caption{{\bf The map projection effect in the power spectrum estimation:} Power spectra estimated with two different projections: the gnomonic projection (left) and the stereographic projection (right). These power spectra (red) are compared to the theoretical power spectrum (black). The right column corresponds to a zoom of the left power spectrum.}
\label{ps_projection}
\end{center}
\end{figure}

In Fig.\ref{ps_projection}, we show the mean power spectra estimated from patches of 20$^{\circ}$x20$^{\circ}$ with a Hann window using two different projections: the gnomonic projection (left) and the stereographic projection (right). The two mean power spectra (red) are compared to the theoretical power spectrum (black). The better result is obtained with the gnomonic projection that will be used as the default projection for power spectrum estimation. However, we have to note that the map projection errors have already been reduced by fixing the size of the field to 20$^{\circ}$x20$^{\circ}$ (see \S \ref{sect_power_method_tiling}).

\subsubsection{Windowing}
\label{sect_power_method_window}
\begin{figure}
\hbox{
\includegraphics[width=5.5cm, height=4.cm]{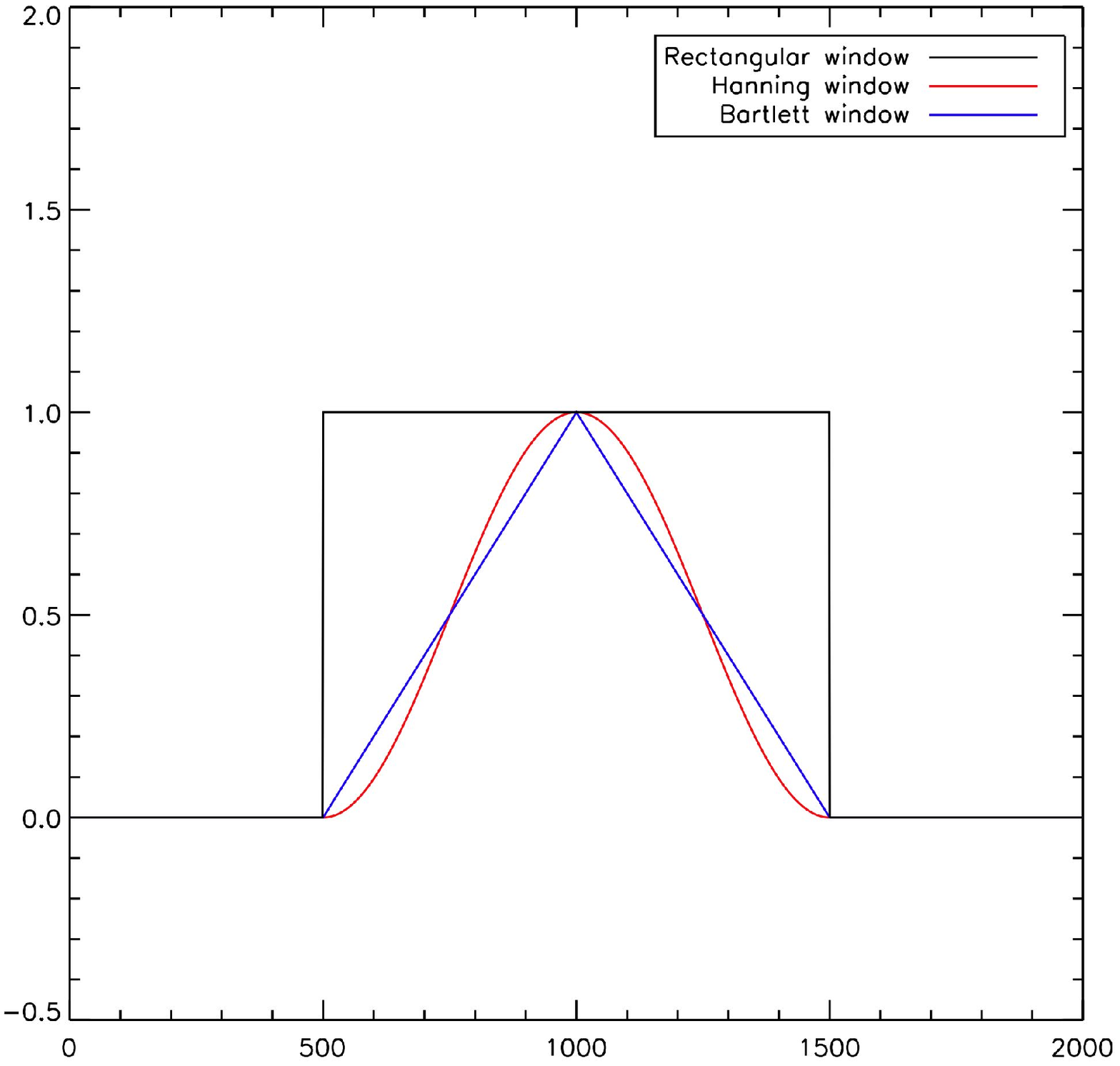}
\includegraphics[width=5.5cm, height=4.cm]{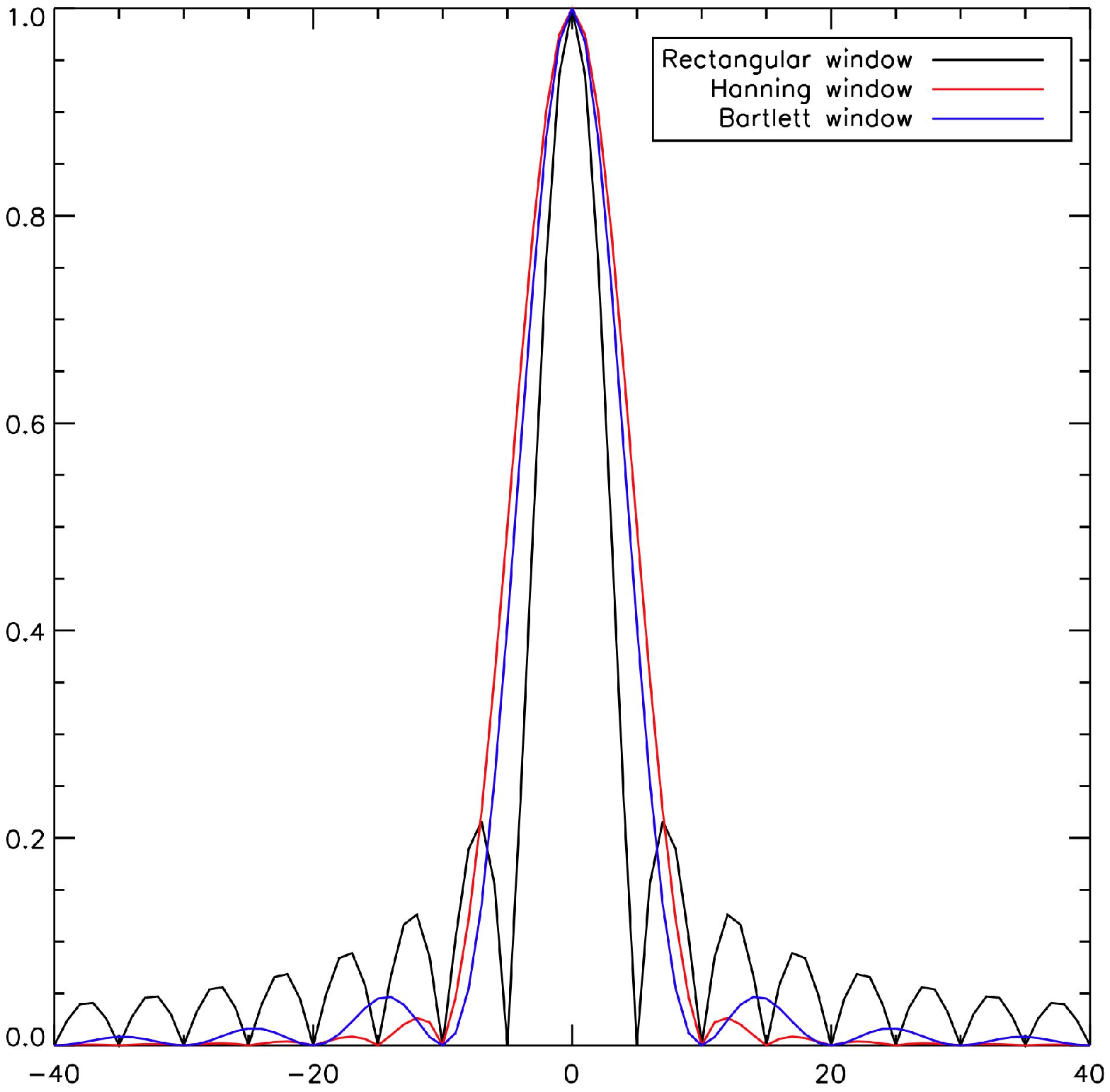}
}
\caption{Window functions commonly used in FFT power spectrum estimation (left) and its frequency response (right). A signal whose frequency is actually located at zero offset leaks into neighboring frequencies with the amplitude shown. The rectangular window which is equivalent to no windowing is the least recommended because of its large side lobes.}
\label{window}
\end{figure}

After the projection, the full-sky CMB is decomposed onto rectangular Cartesian maps of $20^{\circ}$x$20^{\circ}$, but, as already discussed in \S \ref{sect_method_issue}, the analysis of a finite signal affects the frequency analysis. The simplest way to model a patch of a finite size is through the usage of a rectangular window. But, this default window introduces an important spectral leakage. There is a lot of possible other window functions that  can be used to reduce the spectral leakage in the power spectrum estimation. A few of the more common window functions have been compared in this study  (see Fig.\ref{window}):


\begin{itemize}
\item[1.] The Rectangular window that is the default window:
\begin{equation}
h(x, y) = 1.
\label{rectangular}
\end{equation}
\item[2.] The Hann window that has been used in \S \ref{sect_power_method_tiling}:
\begin{equation}
h(x, y) = \cos(\pi x)^2 \cos(\pi y)^2.
\label{hanning}
\end{equation}
\item[3.] The Bartlett window:
\begin{equation}
h(x, y) = 1 - |\frac{x}{2}|-|\frac{y}{2}|.
\label{bartlett}
\end{equation}
\end{itemize}

\begin{figure}[htp!]
\begin{center}
\begin{tabular}{rcc|}
\includegraphics[width=6.2cm, height=4.2cm]{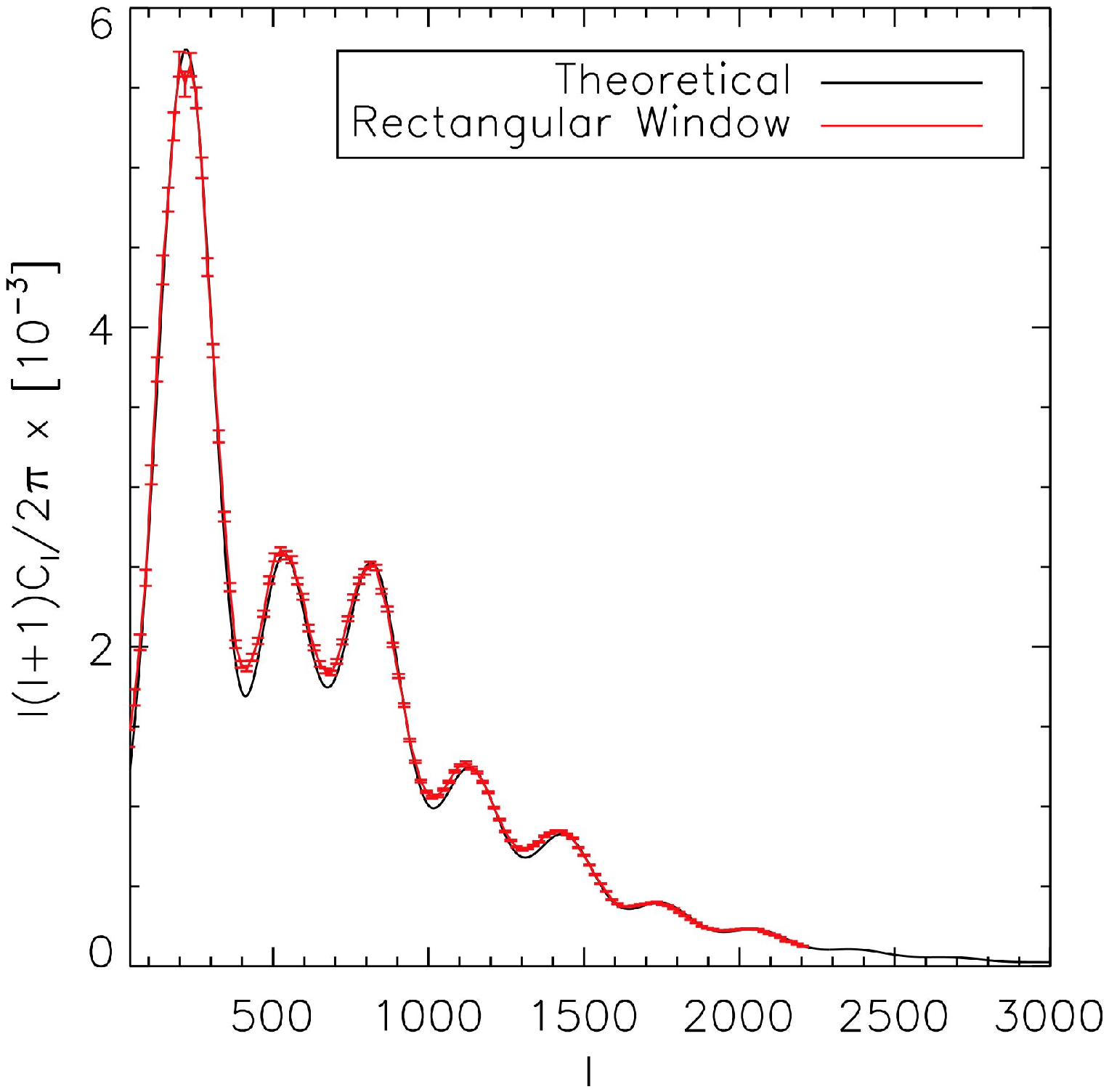}&
~~~\includegraphics[width=5.cm, height=4.2cm]{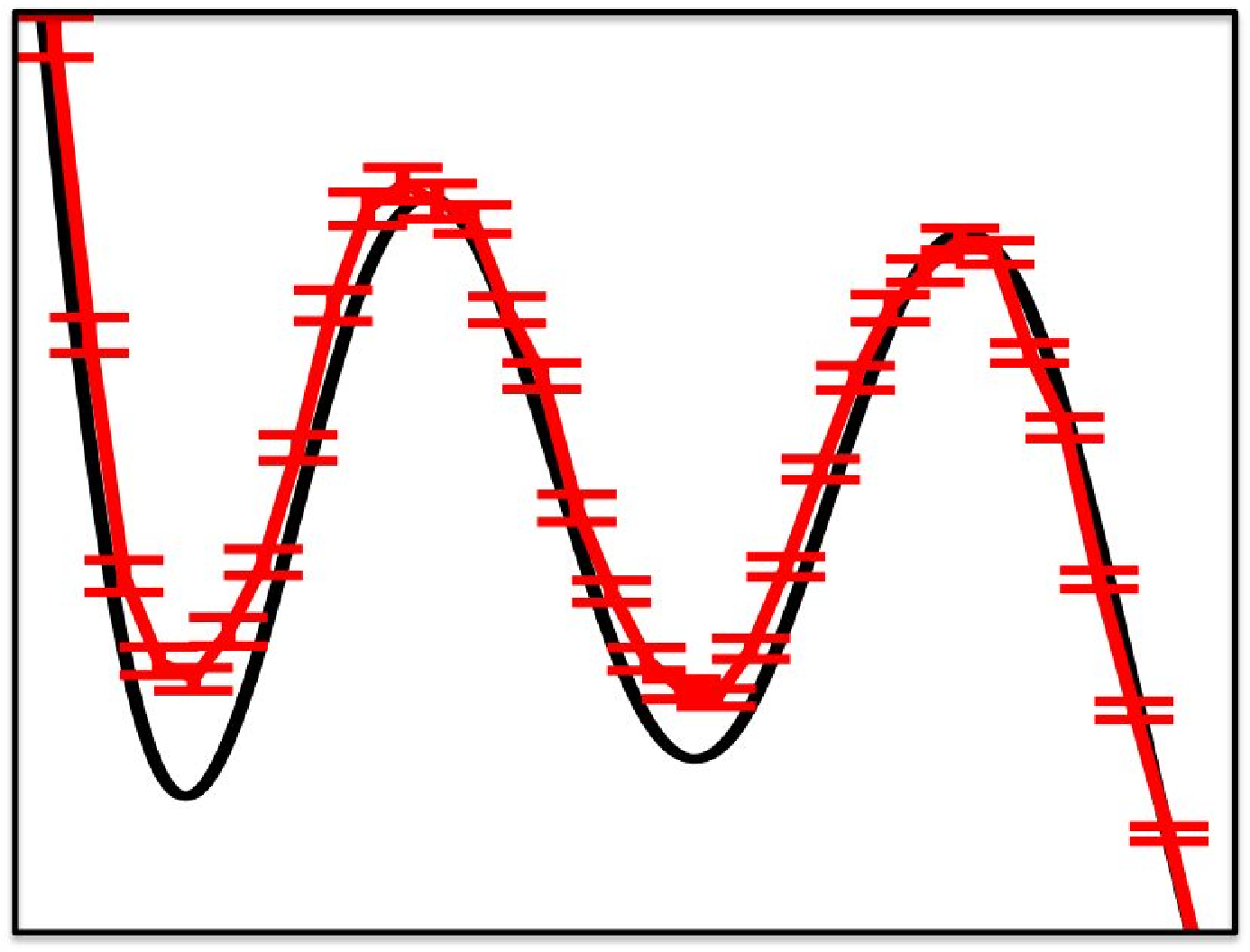}\\
\includegraphics[width=6.2cm, height=4.2cm]{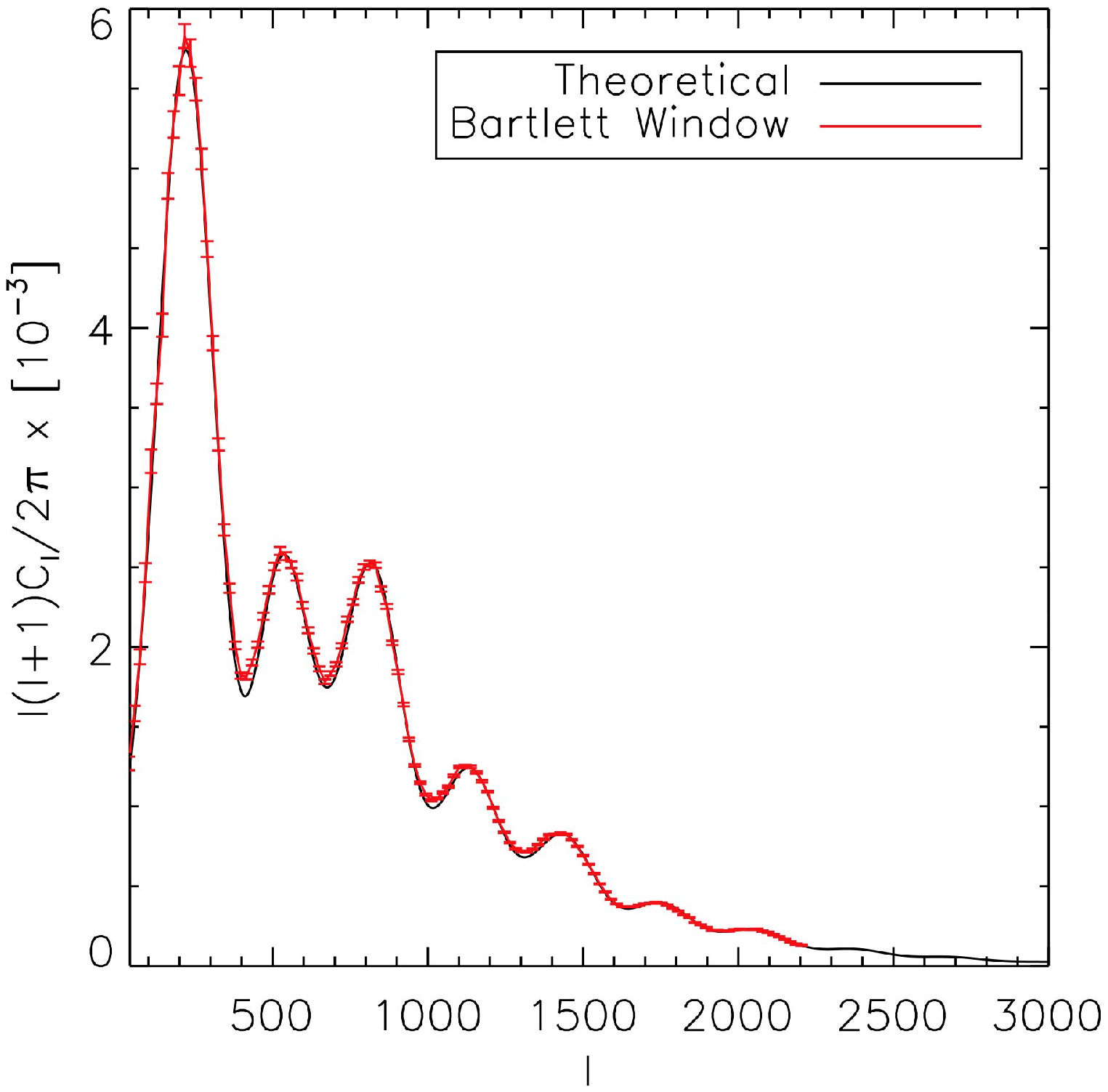}&
~~~\includegraphics[width=5.cm, height=4.2cm]{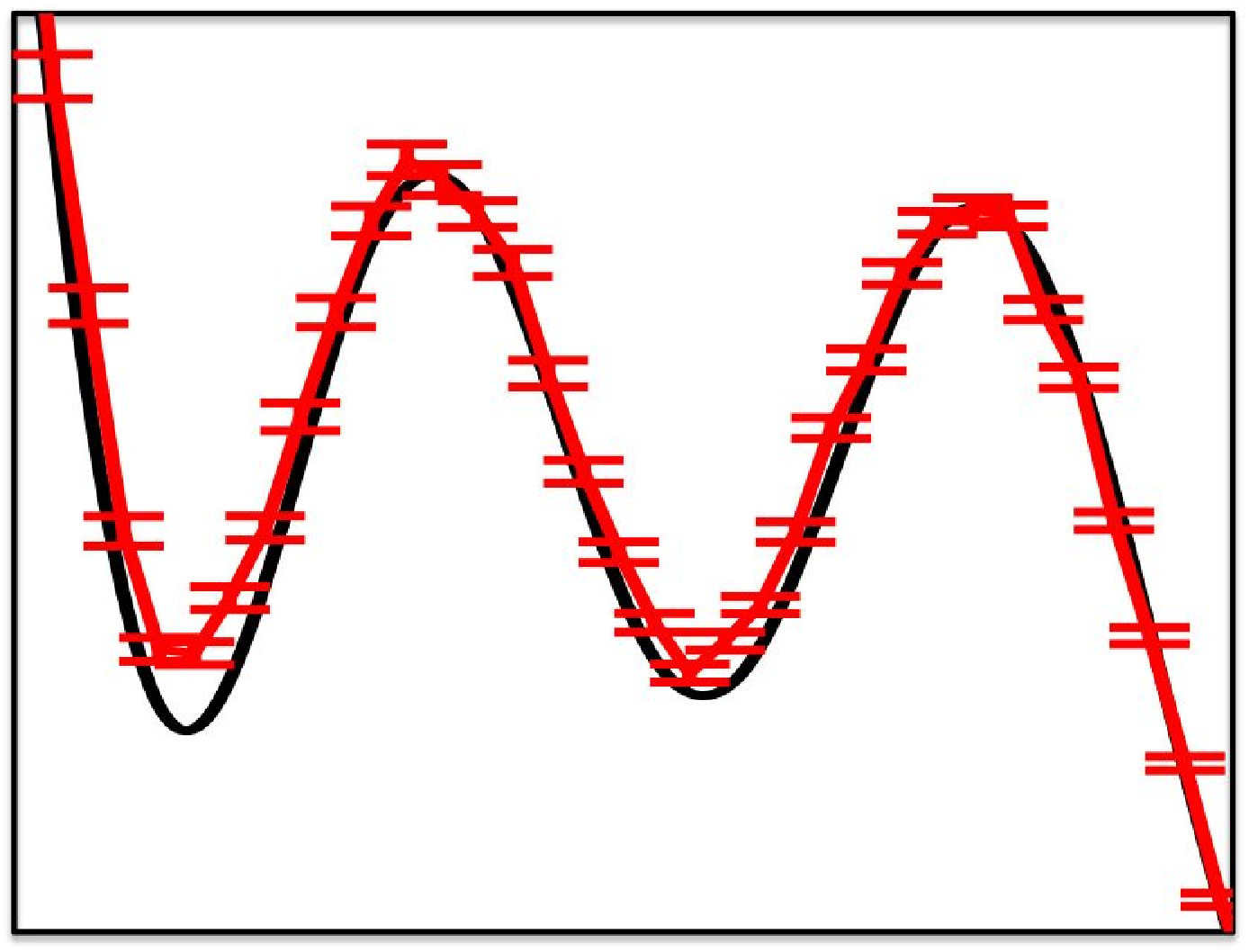}\\
\includegraphics[width=6.2cm, height=4.2cm]{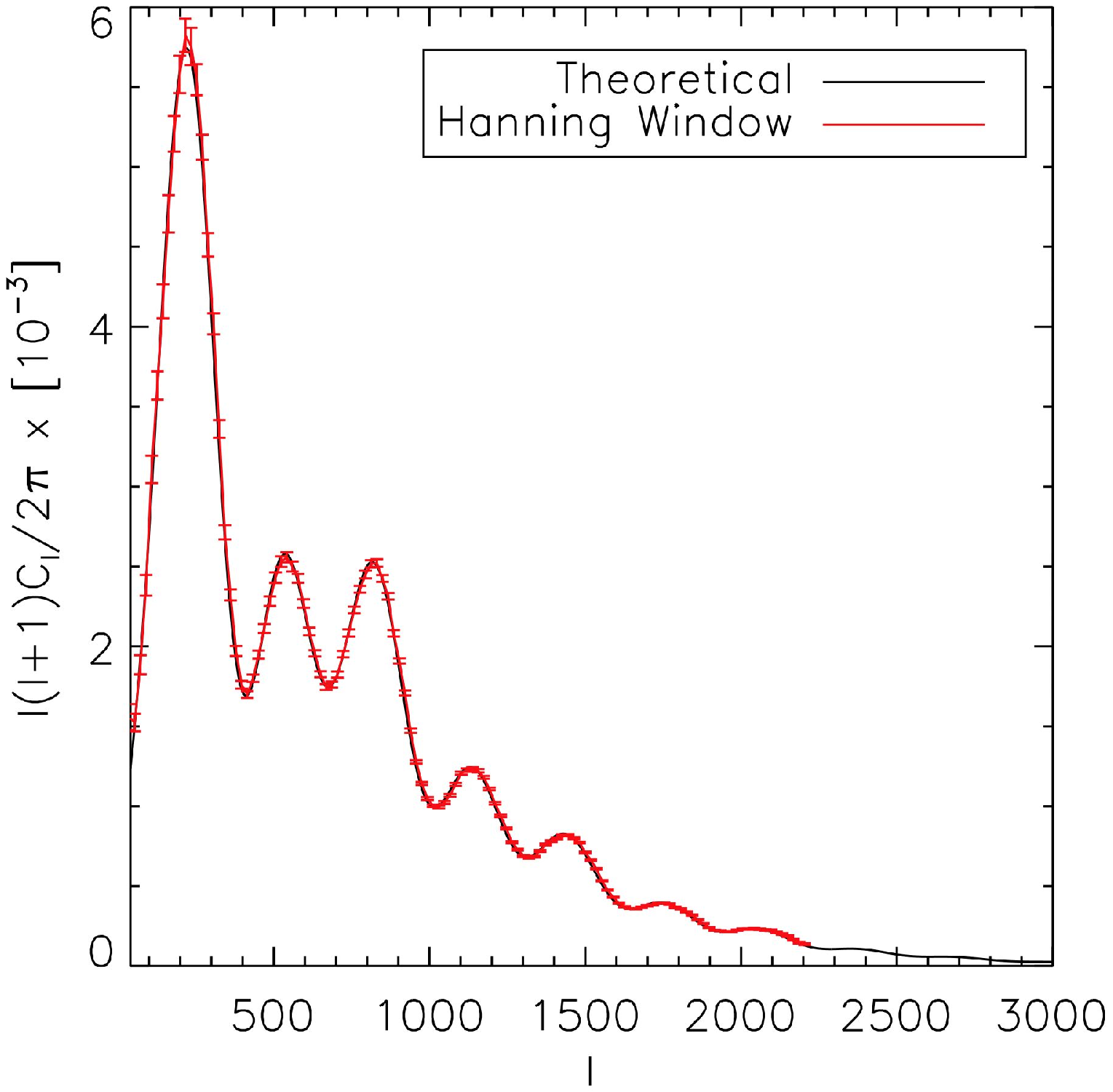}&
~~~\includegraphics[width=5.cm, height=4.2cm]{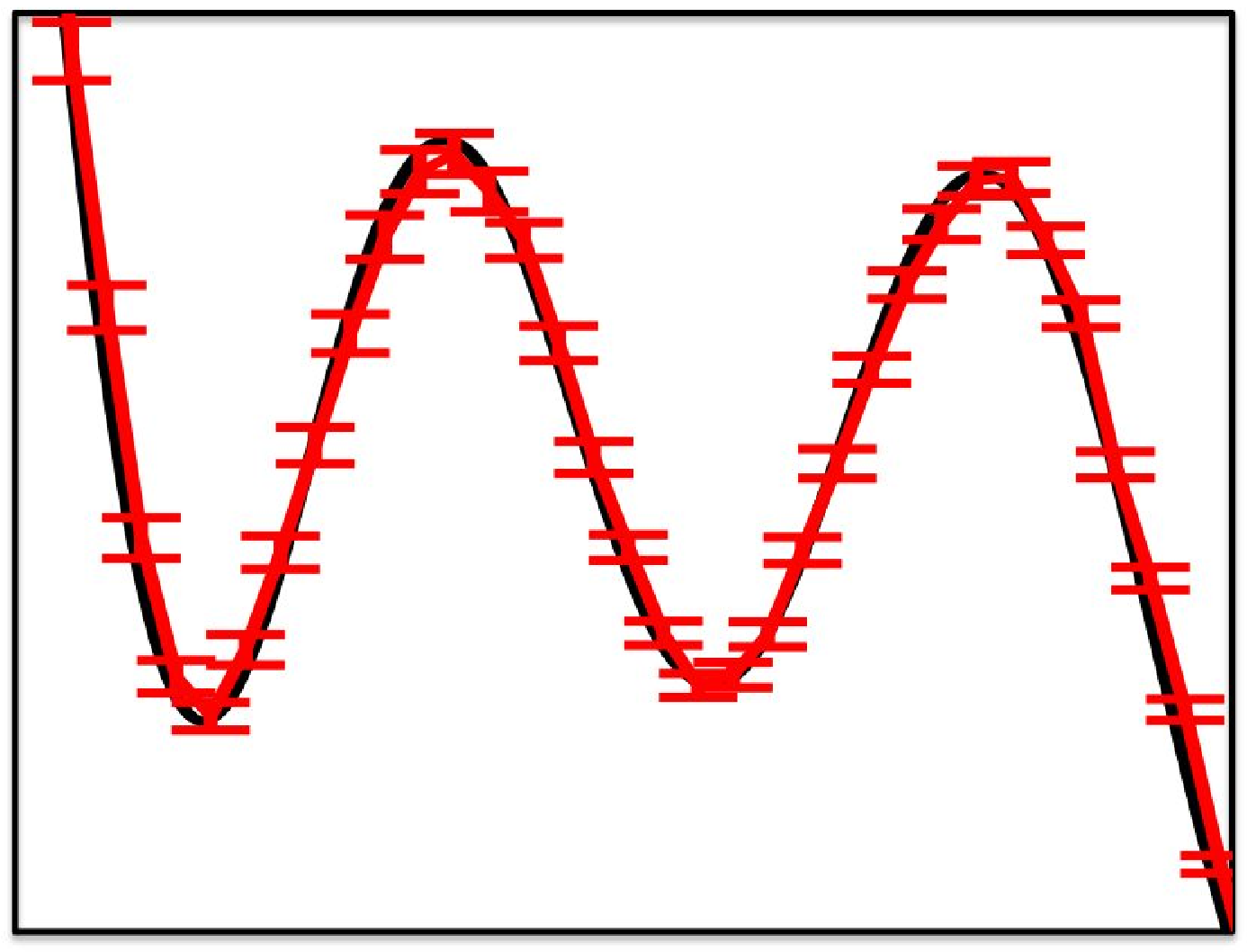}\\
\end{tabular}
\caption{{\bf The window function effect in the power spectrum estimation:} a mean power spectrum (with error bars) is obtained by decomposing the sphere onto patches of $20^{\circ}$x$20^{\circ}$ using a rectangular window (top left), a Bartlett window (middle left) and a Hann Window (bottom left). These power spectra (red) are compared to the theoretical power spectrum (black). The right column corresponds to a zoom in on the left power spectrum.}
\label{ps_window}
\end{center}
\end{figure}


In Fig.\ref{ps_window}, the mean power spectrum has been estimated by decomposing the CMB full-sky onto patches of 20$^{\circ}$x20$^{\circ}$ using the three windows described above. These mean power spectra (in red) have been compared to the theoretical power spectrum that has been used to simulate the CMB full-sky (in black). As expected, a spectral leakage is observed with the default rectangular window due to the finite size of the signal (see the top panel of Fig.\ref{ps_window}). The spectral leakage has been reduced by applying the previously described non-rectangular window functions (see the middle and bottom panel of Fig.\ref{ps_window}). However, the Hann window does the best job (see the bottom panel of Fig.\ref{ps_window}). Indeed, the Hann window is known to produce moderate side lobes (see the bottom panel of Fig.\ref{window}) and to have high frequency resolution which is close to an ideal window function (see \S \ref{sect_method_issue}). Therefore, the Hann window will be used as the default window function for power spectrum estimation.
As said previously, in this paper, we haven't tested the multitaper approach \cite{multitaper:das09} that seems to be optimal to reduce the spectral leakage in the power spectrum estimated from small patches of the sky. This will be done in a future work.

\subsection{Validation of the method with the power spectrum}
In the previous section, we have done an optimization of the method for power spectrum estimation. 
There still room for improvements but is outside the scope of this paper.
The best power spectrum estimation has been obtained by decomposing the sphere onto patches of 20$^{\circ}$x20$^{\circ}$, by projecting the pixels of the patch onto rectangular Cartesian maps using a gnomonic projection and by multiplying each projected map by a Hann window. The power spectrum estimated by this method is now compared with the power spectrum estimated from the spherical harmonics formula (equation \ref{powerspec}).
\label{sect_power_valid}

\begin{figure}[!h]
\hbox{
\includegraphics[width=6.2cm, height=4.2cm]{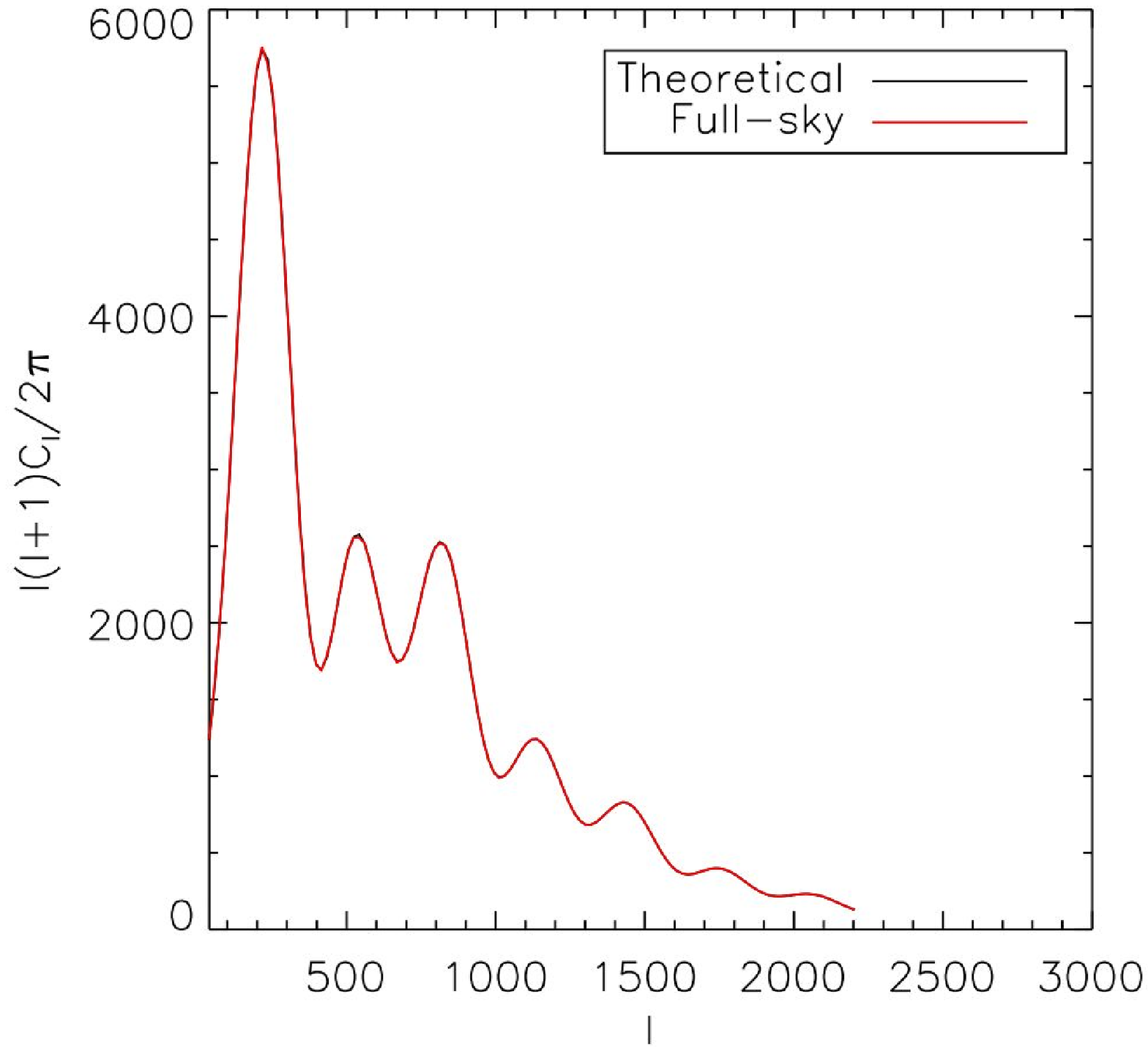}
\includegraphics[width=6.2cm, height=4.2cm]{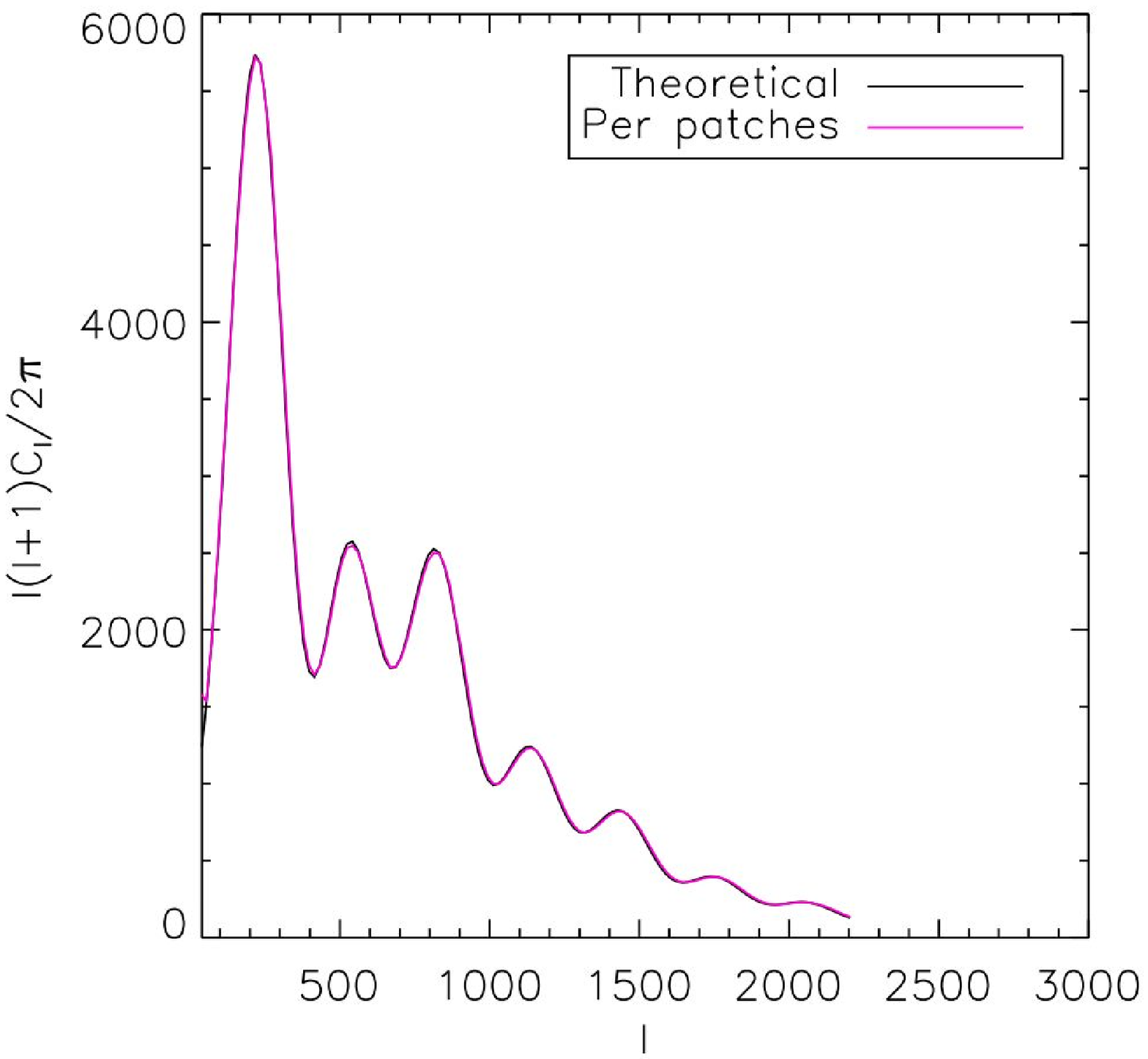}
}
\centerline{\includegraphics[width=6.2cm, height=4.2cm]{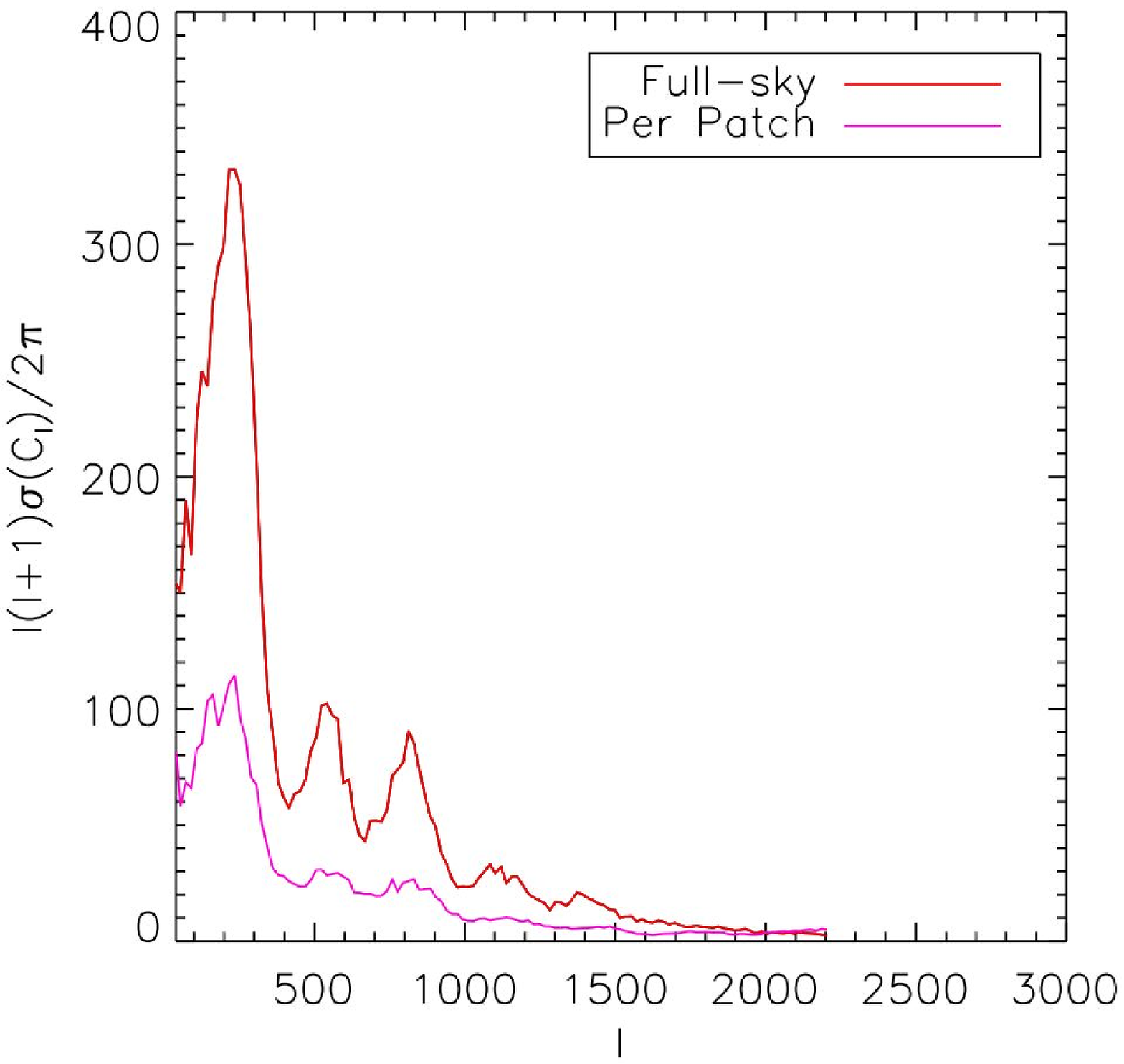}}
\caption{{\bf Mean power spectra estimated from 100 full-sky CMB simulated maps using two different methods:} using the full-sky method based on the spherical harmonic coefficients (top left panel in red) and using the method based on the decomposition of the sphere onto patches (top right panel in pink). These two mean power spectra are compared to the theoretical power spectrum used to produce the simulations (in black). The bias in the mean power spectrum estimated with the full-sky method is about 0.5\% while it is about 3\% with the per patch method. The empirical standard deviation is given for the two methods in the bottom panel.}
\label{ps_comp}
\end{figure}


In Fig.\ref{ps_comp}, we have estimated a mean power spectrum from 100 full-sky CMB simulated maps using the two different methods. On top left, we have the mean power spectrum estimated from spherical harmonics coefficients (in red) and in the top right, the optimized mean power spectrum obtained by the method based on the decomposition of the sphere onto patches (in pink). These two mean power spectra have been compared to the theoretical power spectrum used to do the simulations (in black). The two curves on the left panel lie on top of each other while the two curves in the right panel show a small shift. There still remains a small amount of leakage after windowing. This kind of bias can be corrected \cite{master:hivon02} or reduced using a multitaper approach \cite{multitaper:das09}. The standard deviation is significantly smaller in the per patch method than in the full-sky method (see the bottom panel of Fig.~\ref{ps_comp}), this is because the frequency resolution has been reduced.


It could be thought, the per patch method is not recommended for power spectrum estimation because the decomposition of the sphere onto patches takes some time and there already exists a fast method for a fully CMB power spectrum estimation. However, in practice, we never have access to a full-sky CMB map because of the contamination by residual foregrounds. For this reason, the per patch method can be preferred for some applications.

\subsection{Application of the method}
\label{sect_application}


As mentioned previously, we never have access to a full-sky CMB map. The CMB map obtained by a method of component separation is always partially masked to discard contaminated pixels which introduces problems in power spectrum estimation. About 15-20\% of the most contaminated data is removed mostly in the galactic plane. Several methods exist to overcome the missing data problem, the most common one to be the MASTER method \cite{master:hivon02} that uses apodization windows. The method that has been proposed in this paper is an alternative to methods such as MASTER that try to correct the power spectrum from the effect of missing data. The approach is totally different because the problem of missing data is solved by avoiding the patches with important contaminations on the power spectrum estimation.

Following the same idea, we can imagine to use this method as a diagnosis of the component separation quality. 
An estimator of the power spectrum per latitude can be obtained by averaging the result of the spectral analysis on each patch located at the same latitude (see Fig.~\ref{tiling1}). 
The contamination by residual foregrounds introduces distortions on the power spectrum that should increase close to the galactic plane. The level of contamination per latitude is an indicator of the efficiency of the component separation method. Different component separation methods can be evaluated and compared by this way using simulated data.

\begin{figure}
\hbox{
\includegraphics[width=6.2cm, height=4.2cm]{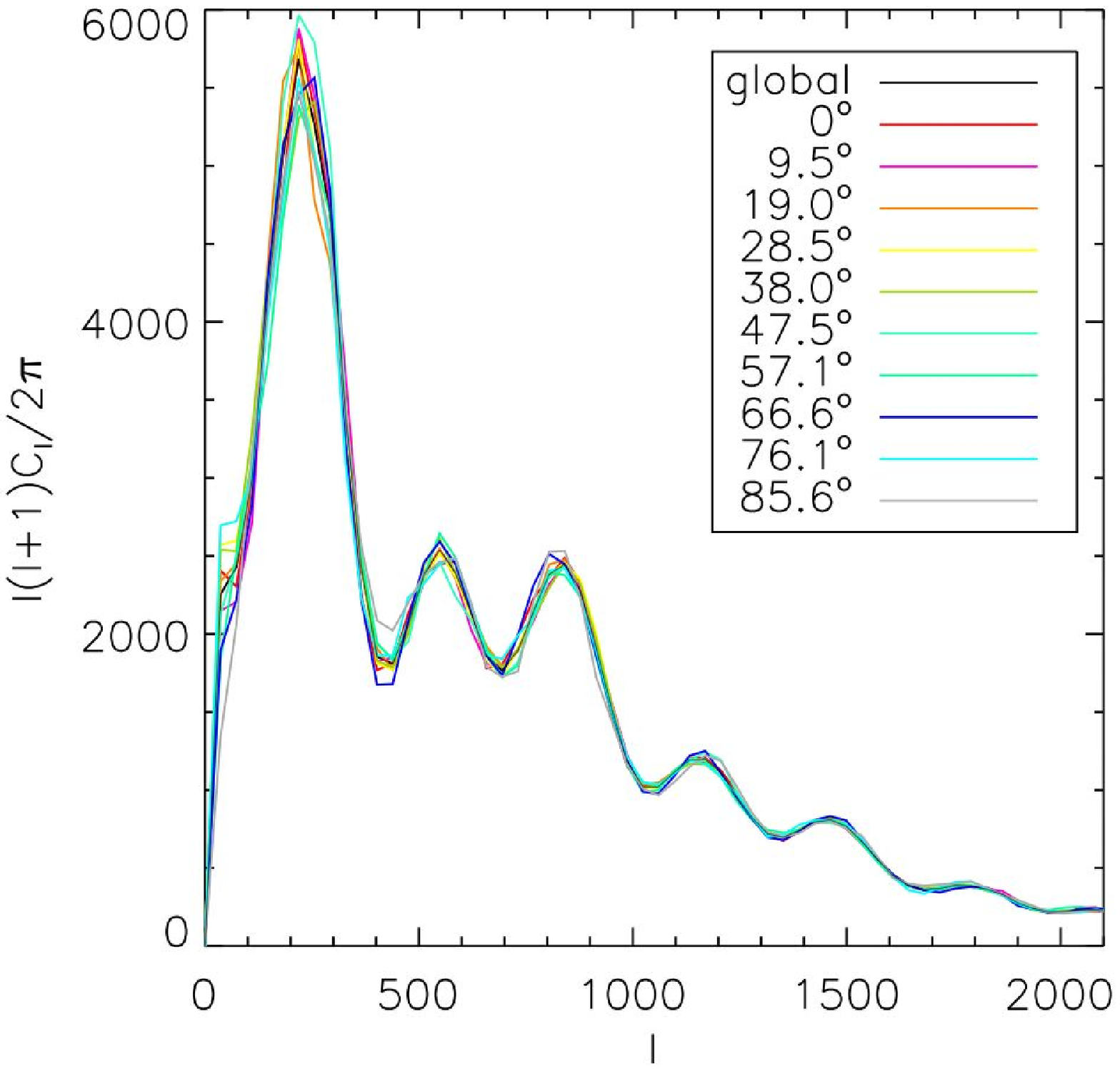}
\includegraphics[width=6.2cm, height=4.2cm]{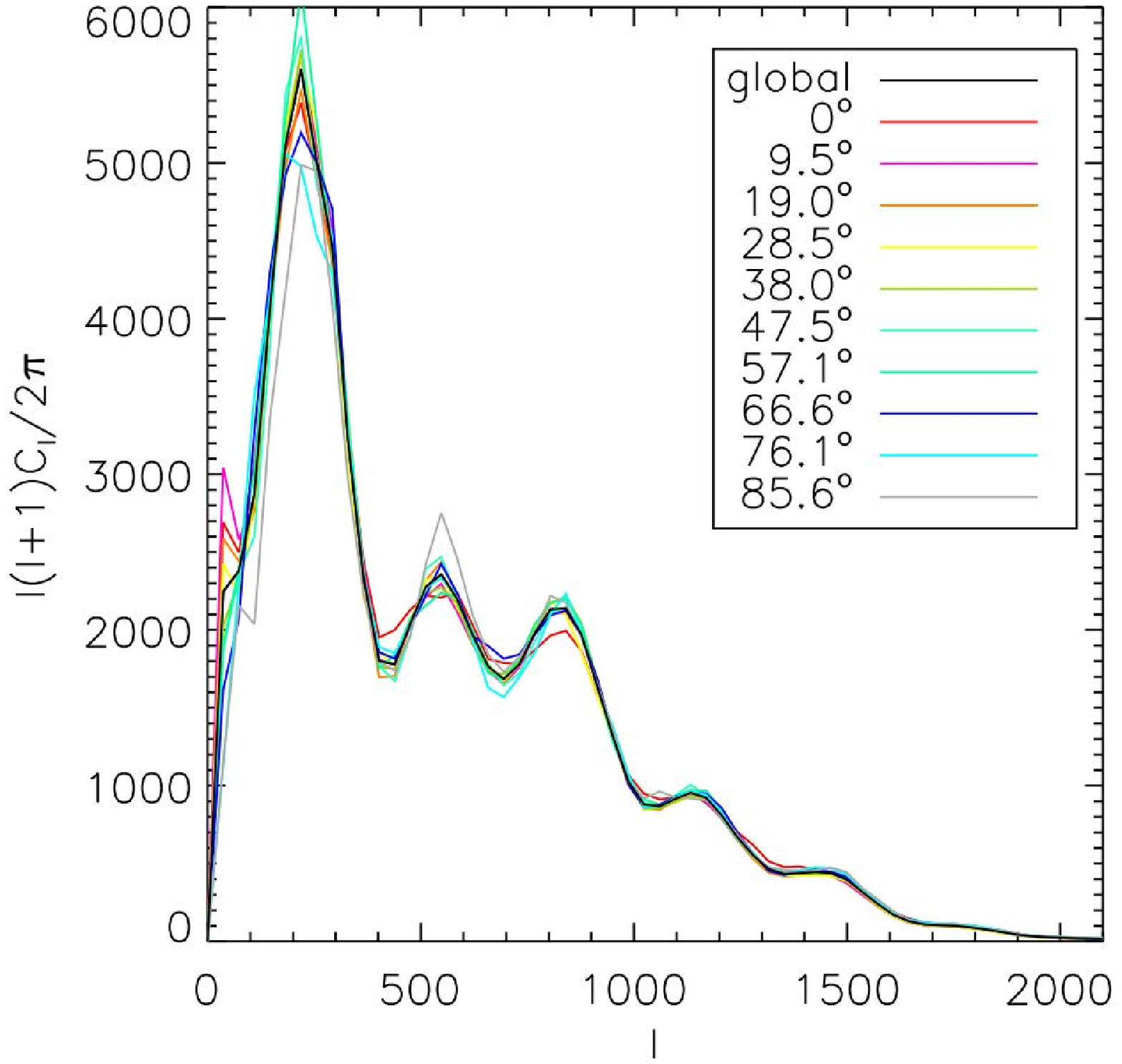}
}
\hbox{
\includegraphics[width=6.2cm, height=4.2cm]{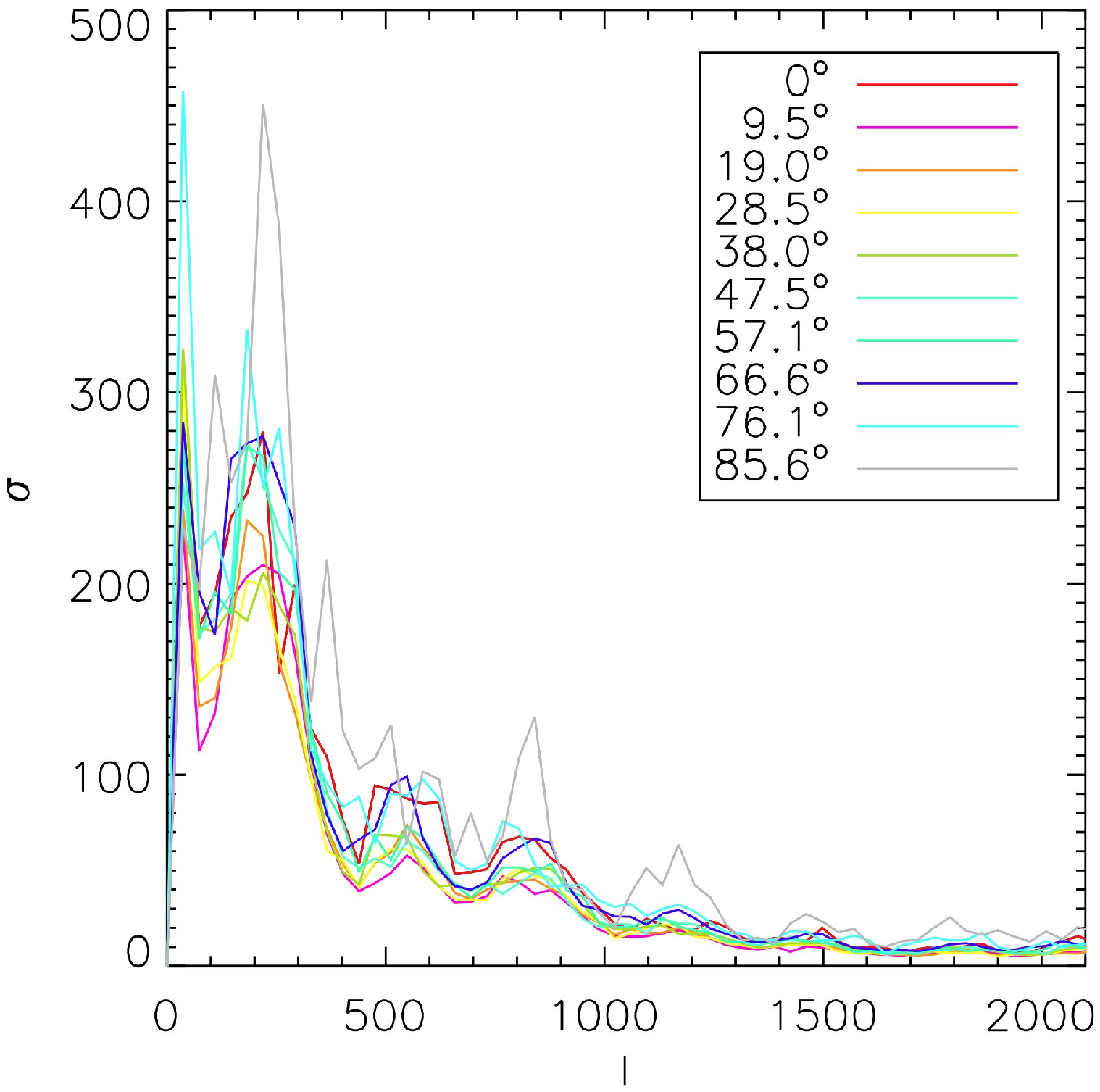}
\includegraphics[width=6.2cm, height=4.2cm]{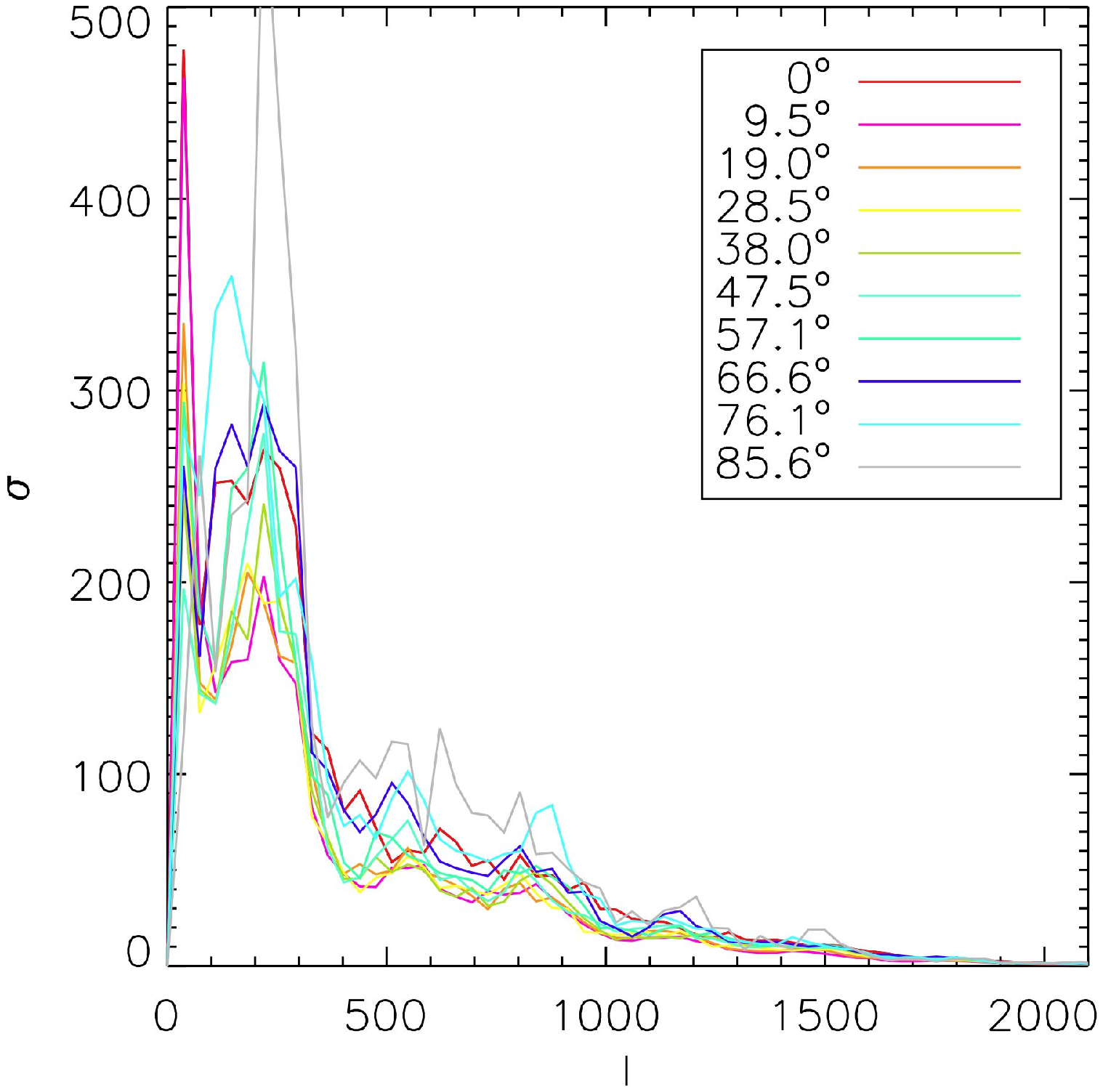}
}
\caption{{\bf Power spectrum per latitude (top) and corresponding errors (bottom)} (patches of $10^{\circ}$x$10^{\circ}$) estimated from a simulated Gaussian CMB map ($n_{side}$ = 2048) without (left) and with (right) contamination of foregrounds. The north and south contributions have been merged.}
\label{ps_latitude}
\end{figure}

Fig.~\ref{ps_latitude} shows the effect of residual foregrounds in the power spectrum estimation.
The left panel shows the power spectra estimated from a simulated CMB map purely Gaussian for different latitudes. The variance is only due to the number of patches per latitude that decreases moving to the poles (see Fig.~\ref{tiling2}). This explains that the statistical variance becomes important for power spectra estimated for latitudes close to the poles (larger than $|70^{\circ}|$).

The right panel of Fig.~\ref{ps_latitude} shows the power spectra per latitude estimated from a simulated CMB map with a significant level of residual foregrounds in the galactic region. This CMB map has been obtained by applying an optimized component separation method to simulations of Planck observations. 
The quality of the component separation method can be evaluated by comparing this result to the previous result. 
As previously, we observe distortions in the power spectrum close to the pole due to the statistical variance (latitude larger than $|70^{\circ}|$). But some important distortions are also present close to the galactic plane (latitude equal to $0^{\circ}$ - red line and $9.5^{\circ}$ - pink line) clearly due to residual foregrounds. This method can be used to select the better component separation method and help to define the region to be masked. Obviously, this only could be accomplished on simulated data for which the true power spectrum is known.

\section{Full bispectrum estimation}
\label{sect_bispec}

The primary goal of this paper is to provide a method for accelerating the calculation of the full bispectrum that cannot be estimated with spherical harmonics in a reasonable time. In the previous section, we have introduced a promising approach that have been tested on the power spectrum estimation. The previous methodology will now be applied to estimate a full bispectrum. Obviously, the method needs now to be optimized for the bispectral estimation.

\subsection{Limitations of the study}
\label{ngsimu}
There will be two limitations to the optimization of this method. The first limitation is the resolution of the non-Gaussian simulations of CMB currently available. The bispectrum of a Gaussian field being null, non-Gaussian CMB simulations have to be used to test the validity of the method. For that purpose, we have used the CMB non-Gaussian simulations of local type provided at the following address: http://planck.mpa-garching.mpg.de/cmb/fnl-simulations. A non-Gaussian CMB temperature map can easily be computed with any desired level of non-Gaussianity ($f^{\rm local}_{\rm NL}$) by linear combination of the $a_{lm}$ provided. More details about the simulations can be found in \cite{simulation:elsner09}. A major problem in our study is that the HEALPix resolution parameter of these simulations is $n_{\rm side}=512$ which is quite low compared to Planck data ($n_{side}=2048$) which has limited our study. But, there exists no other non-Gaussian simulations of CMB publicly available with a better resolution.

The second limitation is that no publicly available codes exist for a theoretical computation of the full bispectrum for a given level of non-Gaussianity, due to its complexity. The comparison with theory is required to be sure that the processing is not introducing important errors in the mean bispectrum. 
An analytic prediction of the bispectrum has been given in \cite{bispec:komatsu02b} for non-Gaussianities of the local-type ($F^{\rm local}_{\rm NL}$) but only for the equilateral configuration. 

In this paper, the comparison with the theoretical bispectrum has been barely done on the equilateral configuration. However,  even focusing on a particular configuration like the equilateral configuration, our method still remains more powerful than $f^{\rm local}_{\rm NL}$ estimation methods because all modes of the equilateral bispectrum are reconstructed compared to a single parameter. 



\subsection{Optimization of the method for bispectrum estimation}
In this section, the optimization of the method for bispectral estimation will be done by comparing the equilateral bispectrum estimated by our method on the CMB non-Gaussian simulations described previously with the analytic prediction given in \cite{bispec:komatsu02b}. However, the low resolution of the non-Gaussian simulations will considerably limit this study. Consequently, more work will have to be done as soon as non-Gaussian simulations with a better resolution will be available.\\

\begin{figure}
\centerline{
\includegraphics[width=9.2cm, height=6.2cm]{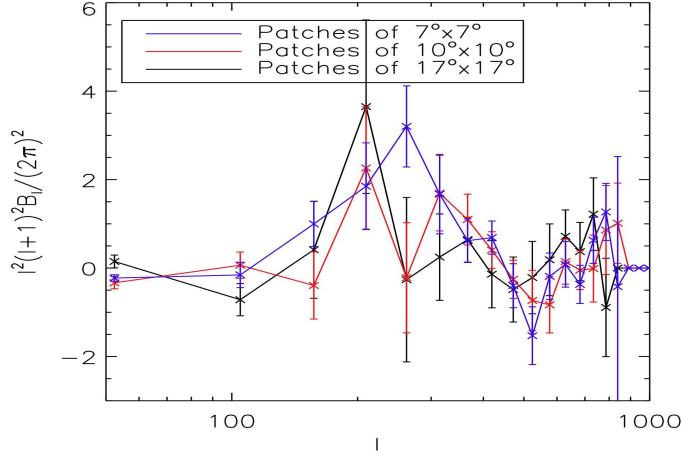}
}
\caption{{\bf The effect of the size of the patches in the bispectrum estimation:} a mean bispectrum has been estimated using three different size of patches ($7^{\circ}$x$7^{\circ}$, $10^{\circ}$x$10^{\circ}$, $17^{\circ}$x$17^{\circ}$) from 100 full-sky non-Gaussian CMB maps ($F^{\rm local}_{\rm NL}=100$).}
\label{bisp_size}
\end{figure}

As for power spectrum estimation, a number of issues has to be solved in order to adapt the previous method to do bispectral estimation. The first issue is the tiling of the sphere. As said previously, for bispectral estimation, the decomposition of the sphere onto rectangular patches located at the HEALPix centers of a lower resolution (see Fig. \ref{tiling2}) is preferred because this decomposition ensures a good repartition of the patches in the sky. But the size of the patches has to be reduced significantly compared to previous decomposition because the bispectrum is very sensitive to the non-Gaussianities introduced by the projection effects. As a result, the gnomonic projection will be preferred to do the projection from the pixels of the patches onto the rectangular Cartesian maps because this projection introduces very little distortions for small patches \citep{projection:goldberg06}. However, the size of the patches has to be fixed in such a way the distortions are reduced.
In Fig.\ref{bisp_size}, the mean bispectrum has been estimated using three different size of field: $7^{\circ}$x$7^{\circ}$, $10^{\circ}$x$10^{\circ}$, $17^{\circ}$x$17^{\circ}$. The pixel projection introduces non-Gaussianities in the projected maps that appears in the bispectrum. For patches of $7^{\circ}$x$7^{\circ}$, the amplitude and the location of the acoustic oscillations are quite well detected. But, the larger the field is, the more important the projection effects are. For patches of $17^{\circ}$x$17^{\circ}$, the CMB non-Gaussianities of the local type almost disappear. There is still more power at the location of the acoustic oscillations but the distortion effects introduce an important noise in the mean bispectrum. This experiment should be redone with simulations with a better resolution.\\

\begin{figure}
\hbox{
\includegraphics[width=6.2cm, height=4.2cm]{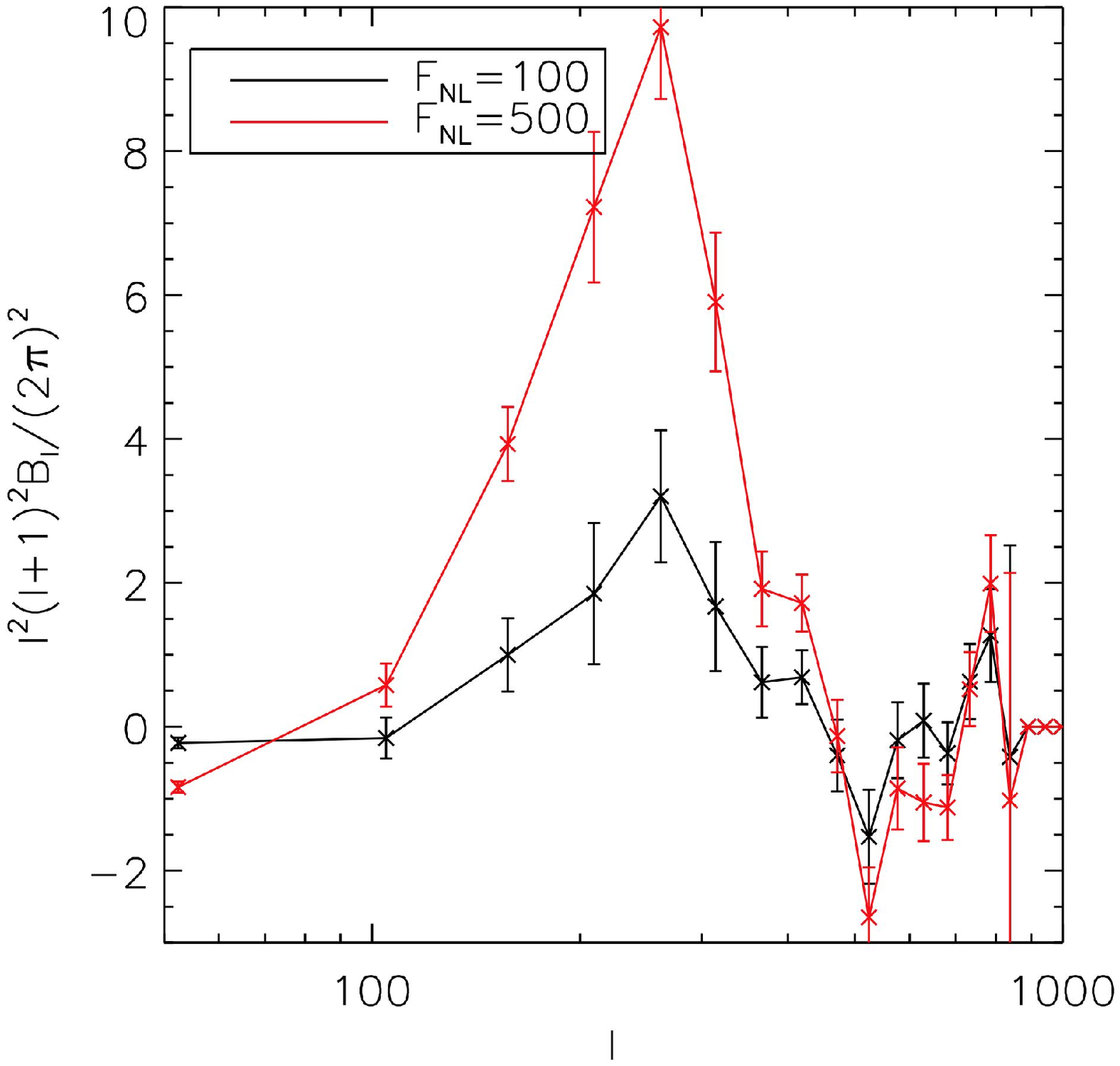}
\includegraphics[width=6.2cm, height=4.2cm]{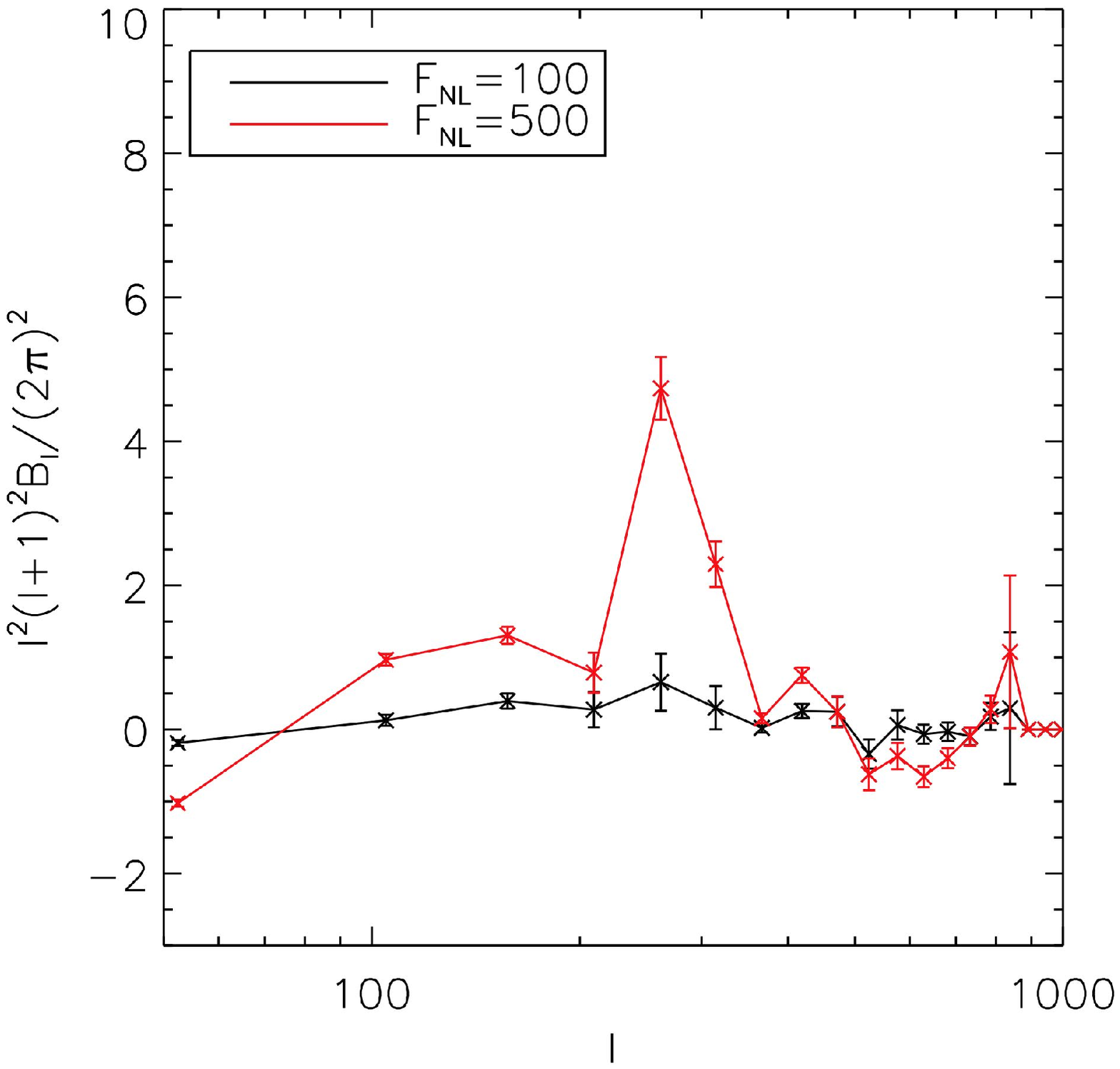}
}
\caption{{\bf The effect of the window function in the bispectrum estimation:} a mean bispectrum has been estimated by decomposing the sphere onto patches of $7^{\circ}$x$7^{\circ}$ using a rectangular window (left) and a Hann window (right) from 100 full-sky non-Gaussian CMB maps with $F^{\rm local}_{\rm NL}=100$ and  $F^{\rm local}_{\rm NL}=500$.}
\label{bisp_window}
\end{figure}

Another issue is the window function. As for spectral estimation, it doesn't exist a perfect window function for bispectral estimation. It should be a trade-off between bispectral resolution and leakage effect. In Fig.~\ref{bisp_window}, we have compared the Hann window to the rectangular window to do bispectral estimation and we obtain a better result with the default rectangular window function (left). Indeed, the amplitude of the acoustic oscillations are better recovered with the default rectangular window. In a future work, a window function dedicated to the CMB data has to be designed to improve the bispectral estimation. Some authors have already tried to address the problem of finding an optimal bispectral window \citep[see][]{window:oroian08,window:wembing02}. \\

\subsection{Preliminary Results}
The best equilateral bispectral estimation has been obtained by decomposing the sphere onto patches of $7^{\circ}$ x  $7^{\circ}$ and by projecting the pixels of the patch onto rectangular Cartesian maps using a gnomonic projection. The code FASTLens \cite{bispec:pires09} publicly available has been used to compute the bispectrum in the Cartesian maps. This study has been limited to the equilateral configuration of the bispectrum. As soon as a code to compute analytical prediction for the full CMB bispectrum will be released, the same study could be extended to all the configurations of the bispectrum. Furthermore, the study will be improved as soon as non-Gaussian simulations of CMB with a better resolution will be available. 

Anyway, the equilateral bispectrum computed (see the left panel of Fig. \ref{bisp_window}) shows the expected acoustic oscillations and the Silk damping predicted by theory for a CMB with non-Gaussianity of the local-type \cite{bispec:komatsu02b}.
Therefore, this preliminary result shows that we can compute an equilateral bispectrum which is in good agreement with the analytical predictions. 



\section{Conclusion}
The goal of this paper is to present a fast way to compute a full-sky CMB bispectrum. It should be pointed out that it is currently too hard to directly measure this full CMB bispectrum. This paper introduces a promising approach for accelerating the estimation of the full bispectrum on the sphere. This method involves the decomposition of the HEALPix map onto small projected Cartesian maps. A mean full bispectrum can then be estimated by combining results from all the projected maps. 

First, this approach has been used to estimate the power spectrum of a full-sky CMB map. A number of optimizations have been done to obtain the best power spectrum estimator. An interesting application of the method in order to test the quality of the CMB component separation on the galactic region has then been presented.

The approach has then be applied to the full bispectrum estimation to accelerate its computation. The approach presented in this paper enables a fast reconstruction of the whole bispectrum directly from the observational data. A number of optimizations have been performed to improve the quality of the bispectral estimation. However, these optimizations has been only tested on the equilateral configuration of the bispectrum because of the lack of analytical predictions for the full bispectrum. Anyway, this study could be easily extended to other configurations as soon as a code to compute analytical predictions for the full CMB bispectrum will be released. Another limitation of  the study comes from the resolution of the non-Gaussian CMB simulations used for the analysis. This is a preliminary result that should be improved as earlier as is feasible. 

However, the equilateral bispectrum that is computed from the non-Gaussian CMB simulations using this approach is in very good agreement with the analytical predictions.  Indeed, the features expected by the theory are present despite the poor resolution of the estimated bispectrum. Thus, this approach appears very promising to constrain the CMB non-Gaussianity in a model-independent way.

\section*{Acknowledgment}
This work has been supported by the European Research Council grant SparseAstro (ERC-228261). Some of the results in this paper have been derived using the HEALPix package (Gorski, Hivon, and Wandelt 1999). We wish to thank F. Sureau, P. Paykari and Y. Moudden for useful discussions and comments.


\end{document}